# ATP synthase: evolution, energetics, and membrane interactions

Jasmine A Nirody[1,2,*], Itay Budin[3], and Padmini Rangamani[4*]

[1]Center for Studies in Physics and Biology, The Rockefeller University, New York, NY 10065
[2]All Souls College, University of Oxford, Oxford, UK OX14AL
[3]Department of Chemistry and Biochemistry, University of California San Diego, La Jolla CA 92093-0375
[4]Department of Mechanical and Aerospace Engineering, University of California San Diego, La Jolla CA 92093-0411

*Correspondence: jnirody@rockefeller.edu; prangamani@ucsd.edu

## Abstract

The synthesis of ATP, life's 'universal energy currency', is the most prevalent chemical reaction in biological systems, and is responsible for fueling nearly all cellular processes, from nerve impulse propagation to DNA synthesis. ATP synthases, the family of enzymes that carry out this endless task, are nearly as ubiquitous as the energy-laden molecule they are responsible for making. The F-type ATP synthase (F-ATPase) is found in every domain of life, and is believed to predate the divergence of these lineages over 1.5 billion years ago. These enzymes have therefore facilitated the survival of organisms in a wide range of habitats, ranging from the deep-sea thermal vents to the human intestine. In this review, we present an overview of the current knowledge of the structure and function of F-type ATPases, highlighting several adaptations that have been characterized across taxa. We emphasize the importance of studying these features within the context of the enzyme's particular lipid environment: Just as the interactions between an organism and its physical environment shape its evolutionary trajectory, ATPases are impacted by the membranes within which they reside. We argue that a comprehensive understanding of the structure, function, and evolution of membrane proteins -- including ATP synthase -- requires such an integrative approach.

## Introduction

The use of ATP as a source of chemical energy to drive metabolic activity is ubiquitous and shared among all known cellular life forms. The energy stored in ATP's phosphoanhydride bond is used to power a wide range of processes including muscle contraction, cell motility, nerve impulse propagation, and DNA synthesis, among many others. This impressive task list has earned the molecule the title of the 'universal energy currency'. As might be expected, the turnover of ATP is also remarkable: the human body uses about its weight in ATP daily (Dimroth et al., 2006), and cells must continually regenerate ATP from the products of its hydrolysis (ADP and phosphate) to keep up with this demand. ATP synthases, the enzymes that perform this tireless job are nearly as universal as the currency they are responsible for minting.

The ATP synthase family of enzymes comprises three members: A-, V-, and F-type ATPases/ATP synthases (Kagawa and Racker, 1966; Penefsky et al., 1960). *In vitro*, all three are reversible: they can both use the energy of ATP hydrolysis to move cations across ion-impermeable membranes, as well as use the energy stored in the transmembrane ion gradient to synthesize ATP. *In vivo*, the V-type acts as an ATPase (Forgac, 2007). The A- and F-types can work in either direction in prokaryotes (Müller and Grüber, 2003), though F-type ATPases act as purely an ATP synthase in eukaryotic mitochondria and chloroplasts (Kühlbrandt, 2019).

A-type ATPases exist in *a*rchaea (Müller and Grüber, 2003), although they are also utilized by a small number of bacterial species, which appear to have obtained them via horizontal gene transfer (Lapierre et al., 2006). V-type ATPases belong to the eukaryotes and sit primarily in the membranes of *v*acuoles (Forgac, 2007). The final class, the F-type ATPases, is the only one with members found in every branch of the tree of life: present in both bacterial and mitochondrial membranes (von Ballmoos et al., 2009), they were originally characterized as *c*oupling *f*actors (CF-ATPases). Later, when they were discovered in chloroplasts, the 'c' was dropped to avoid confusion (Kühlbrandt, 2019). The reach of F-type ATPases then extended to Archaea when a subclass (the *n*ovel, or N-type) was identified in some archaeal species – once again, thought to be a consequence of a lateral gene transfer event (Dibrova et al., 2010).

Due to their prevalence in organisms ranging from archaea to animals, we focus on diversification in F-type ATPases/ATP synthases. For ease of exposition, we will refer to this family of enzymes as ATPases in the remainder of this text. We begin with a brief overview of shared structural, functional, and phylogenetic features between the classes of ATPases. For more detailed reviews on the other members of the ATPase family, we refer interested readers to the following excellent reviews (Forgac 2007 for V-type ATPases; Grüber et al., 2014 for A-type ATPases).

Structurally, all rotary ATPases share basic similarities: each is composed of a soluble domain (termed A1, V1, or F1, for the A-, V-, and F-type, respectively) coupled to a membrane-embedded domain (A0, V0, or F0) by a connecting stalk (Kühlbrandt 2019). The former contains the ATP hydrolyzing/synthesizing domain, while the latter facilitates ion translocation across the membrane. Each domain on its own is an independent stepping motor; this modularity has, for instance, been demonstrated by experiments showing that detached F1 and A1 retain rotary movement driven by ATP (Imamura et al., 2003; Noji et al., 1997). The membrane domain is often described in terms analogous to macroscopic motors: the rotary portion of the motor is designated the 'rotor', while the static portion is fixed to the soluble domain by the motor's 'stator'. In the 'ion pump' or ATP hydrolysis mode, the soluble domain sequentially drives rotation of an 'axle' that transmits energy to the membrane-embedded component; when in the 'power' or ATP synthesis mode, ion movement through the membrane domain drives axle rotation (Muench et al., 2011).

Based on sequence and subunit composition, A-type ATPases are more closely related to V-type ATPases than to F-type ATPases. Functionally, however, the A- and F-types show more similarities -- for instance, the ability to synthesize ATP *in vivo* (Ihara et al., 1992; Iwabe et al., 1989; Müller et al., 1999; Nelson, 19992; Schäfer et al., 1999). These similarities in sequence, structure, and function point to the existence of a shared common ancestor, which underwent a series of changes to arrive at the three modern ATPase classes (Imamura et al., 2003; Zimniak et al., 1988). Due to the conservation of their basic structure and function across biological kingdoms, the ATP synthases are an intriguing framework within which to study several major events in the evolution of life, from its origin and the rooting of the tree of life (Mulkidjanian et al., 2009) to the prokaryote-eukaryote transition (Hilario and Gogarten, 1998), among others. There have been several excellent pieces of work on the diversity of ATPases, which utilize sequence data, high-resolution structures, and functional assays (von Ballmoos et al., 2007; Muench et al., 2011; Müller and Grüber, 2002).

However, the dynamics of diversification in the ATPase lineage must take into account the 'ecology' of these enzymes: just as the interactions between an organism and its physical environment shapes its evolutionary trajectory, ATPases do not perform their duties in a vacuum. Indeed, there is evidence of coevolution between biological membranes and membrane proteins, suggesting that a deeper understanding of the evolutionary history of the ATPases may provide key insights into the prototypic organism from which sprouted the tree of life (see, e.g., Mulkidjanian et al., 2009). Studies have shown how ATP synthase dimers can shape mitochondrial cristae (Seelert and Dencher, 2011); dimers have been observed in a small fraction of chloroplast ATP synthases as well, though likely through an independent mechanism (Rexroth et al., 2004).

There are myriad adaptations that have allowed ATPases to drive the survival of organisms in such a wide range of environments – many that have yet to be fully characterized, and too many to squeeze into even the most exhaustive review. Here, we begin with a comparative overview of the structure and function of F-type ATPases in mitochondria, chloroplasts, and prokaryotes (here, we group together the bacteria and archaeal enzymes due to their purported common origin). We then discuss several adaptations that have been observed across species, either via a shared ancestor or through convergent evolution.

## F-ATPase structure and function across biological domains

Like their siblings, F-ATPases are 'dual motors', composed of two modular units, the cytoplasmic F1 and the membrane-embedded Fo. Prokaryotic and chloroplast ATPases are relatively simple in structure, while mitochondrial enzymes are more complex and variable between species and cell types; a comparison of these structures is provided in Figure 1.

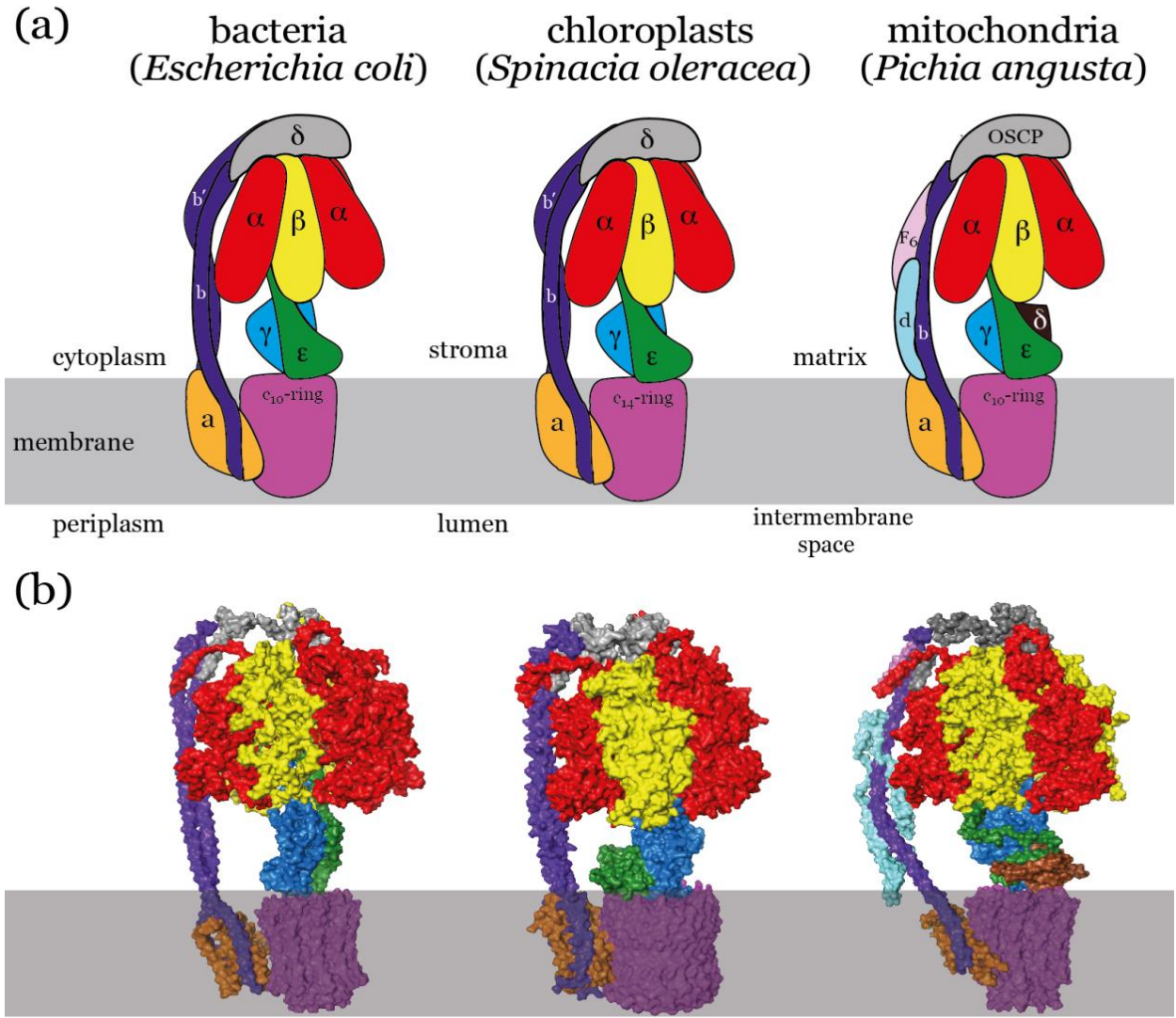

**Figure 1** Comparative *(a)* schematic and *(b)* atomic models of F1Fo motors from bacteria, chloroplasts, and mitochondria. Map resolution are as follows: 6.9 Å for *Escherichia coli* ATP synthase (Sobti et al., 2016); 2.9 Å/3.4 Å for *Spinacia oleracea* (spinach) chloroplast F1/Fo, respectively (Hahn et al., 2018); 7.0 Å for *Pichia Angusta* mitochondrial ATP synthase (Vinothkumar et al., 2016). Subunits are color coded to show homologous domains across species: *α* (red); *β* (yellow), *δ* (grey in bacteria and chloroplasts, brown in mitochondria); OSCP (grey in mitochondria); *b/b'* (dark purple); *c*-ring (light purple); *F6* (pink); *γ* (blue); *ε* (green); *d* (light blue).

The catalytic F1 'head', the site of ATP synthesis, is a mushroom-shaped extension that pokes approximately 11 nm into the cytoplasm. The F1 subunit comprises three *α*- and three *β*-subunits arranged around a central *γ*-subunit. As F1's membrane-bound counterpart, the Fo sector is responsible for torque generation during synthesis ('power mode'), and for ion translocation in the reverse direction ('ion pump mode'). The membrane-bound motor contains the *a*-subunit and a 'rotor ring' of 8 to 17 *c*-subunits, depending on the species (Meier et al., 2005). The geometry of the side chains of the *c*-ring determine the ion-selectivity in bacterial motors (Dimroth et al., 1999; Krah et al., 2010) – that is, whether the motor runs on the proton motive force (*pmf*) or the sodium motive force (*smf*).

The F1 and F0 sectors are connected by two stalks: one central, composed of subunits $\gamma$, $\varepsilon$, with the additional $\delta$ in mitochondria; and one peripheral, composed of the *b*-subunits, and with the additional *d* and *F6* in mitochondria (Wilkens and Capaldi, 1998). The central stalk is responsible for transferring torque generated by F0 to the catalytic F1 head, shifting it between the three confirmations shown in Figure 2. The $\varepsilon$-subunit has also been shown to have some regulatory functions in bacteria and chloroplasts (Feniouk et al., 2006; Gibbons et al., 2000; Laget and Smith, 1979; Richter et al., 1984; Smith et al., 1975; Stock et al., 2000); additionally, in chloroplasts, the $\gamma$-subunit acts as a regulator to switch off nighttime ATPase activity in chloroplasts (Kohzuma et al., 2013). Mitochondrial ATPases are regulated by a separate small inhibitory protein IF1 (Bason et al., 2014; Gledhill et al., 2007; Pullman and Monroy, 1963). The peripheral stalk is connected to F1 by the $\delta$-subunit in chloroplasts (unrelated to the $\delta$-subunit in the mitochondrial central stalk, except through a confusing naming convention!) and by OSCP in mitochondria. Sequence studies indicate that the $\delta$-subunit and OSCP are nearly the same protein, indicating that the peripheral stalk has remained unchanged for over 1.5 billion years (Hohmann-Marriott and Blankenship, 2011; Rand et al., 2004).

The catalytic sites in the F1 sector, where the nucleotides bind, are contained mostly in the *β*-subunits, with the *α*-subunits contributing an essential arginine residue. These sites go from being *e*mpty ('*β*E') to being occupied by Mg-*ADP* and phosphate ('*β*DP'). Driven by the current flow through F0, F1 can catalyze these ligands to form Mg-*ATP* at the third catalytic site ('*β*TP') (Gibbons et al., 2000; Stock et al., 2000). When operating in the other direction ('ion pump mode'), the catalysis of Mg-ATP drives rotation of the *c*-unit via the central stalk, facilitating the transmembrane flow of ions (Junge at al., 1997; Cherepanov et al., 1999; Feniouk et al., 2000; Mulkidjanian et al., 2006).

The three rotational states in the F1 sector were resolved through high-resolution cryo-EM studies in bacteria (in *E. coli*, Sobti et al., 2016; and *Bacillus* PS3 Guo et al., 2019) and chloroplasts (in spinach; Hahn et al., 2018). Using fluorescently labeled bacterial F1, the $\gamma$-subunit was shown to rotate at over 130Hz, indicating that bacterial F-ATPases can synthesize (or hydrolyze) approximately 400 molecules of ATP every second (Fischer et al., 1994; Yasuda et al., 2001)! The OCSP region has been shown to be responsible for the flexible coupling between the two sectors, consolidating F1's three discrete catalytic steps with F0's more continuous rotation (Murphy et al., 2019). A schematic of this cycle, termed the binding change mechanism (Boyer 1997; Boyer at al., 1977), is shown in Figure 3. Several theoretical models have been posed, at varying levels of abstraction, to describe this mechanism (Elston et al., 1998; Mukherjee and Warshel, 2012; Oster et al, 2000; Oster and Wang, 2000; Sun et al., 2004; Wang and Oster, 1998; Xing et al, 2005).

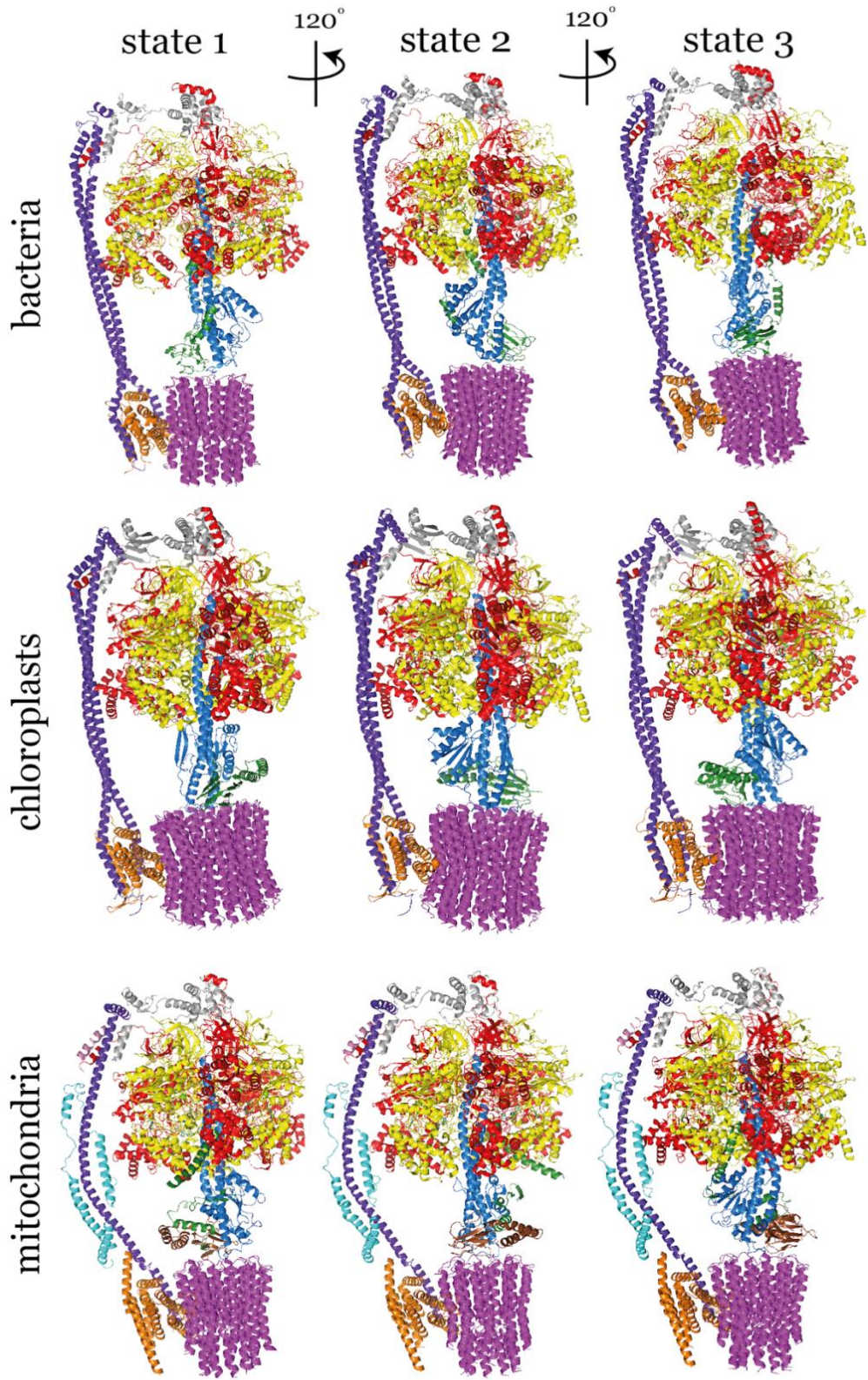

**Figure 2.** Three primary rotary states of intact F1Fo ATPases in bacteria (*E. coli*), chloroplasts (*S. oleraca)*, and mitochondria (*P. angusta*), separated by an approximately 120° rotation. States are listed from most to least populated; the most populated states for each organism (conformation 1) are as shown in Figure 1.

## Fueling up: selective ion binding and transport

The vast majority of extant metabolic strategies are believed to have been in stasis after their emergence in the first billion years after the origin of life on earth (Banfield and Marshall, 2000; Battistuzzi et al., 2004; Knoll and Bauld, 1989). Despite this concentrated time scale, organisms have developed many ways to make and use energy; including aerobic and anaerobic respiration, methanogenesis, sulfur and iron reduction or oxidation, and several others (Boucher et al., 2003; Castresana, 2001; Koumandou and Kossida, 2014; Reysenbach and Shock, 2002).

The diversity of these metabolic modes is perhaps most obvious in the prokaryotes. Interestingly, these pathways exhibit a 'patchy distribution' across the prokaryotic phylogeny, indicating over 25 independent origins or over a dozen horizontal gene transfer (HGT) events. However, there is no such evidence of such 'patchiness' in the universal enzyme central to all these pathways: F-ATPase. A phylogenetic analysis of all the subunits of F-ATPase showed that enzyme sequence and structure did not cluster with bioenergetic mode. This analysis confirms the widespread utilization, and thus ancient origin, of ATP synthase. Importantly, it also points to its robustness: F-ATPase, showing few specific modifications, is able to function as part of a wide variety of electron transport chains, and within a wide variety of cell types (Koumandou and Kossida, 2014).

One of the most basic questions one can ask about an ion driven enzyme is about its choice of fuel. For ATP synthases, however, this question has also proven to be one of the most puzzling. Environmental factors, such as temperature or pH, have not been found to consistently determine several features of ATP synthases, including ion selectivity. Under most physiological conditions, the $Na_+$ gradient far surpasses that of $H_+$ — often by a millionfold (e.g., in the mitochondria) or more (e.g., in alkaline environments). But despite this environmental bias, nearly all F-ATPases (and their V-type siblings) use proton flow as their power source. Structurally, membrane proteins within the same family driven by *pmf* or *smf* are largely conserved (Ethayathulla et al., 2014; Meier et al., 2005; Pogoryelov et al., 2009; Ressl et al., 2009; Yamashita et al. 2005), indicating that 'switching' between these two energy sources occurs via localized changes in amino-acid composition rather than any dramatic structural modifications (Leone et al., 2015). It is somewhat surprising, then, that the strong $H_+$ selectivity required to counteract the environmental sodium excess is achievable through such minor modifications.

Ion specificity in ATP synthases is the domain of the membrane embedded Fo sector. The *a*-subunit provides an aqueous pathway for ion transport onto binding sites along the ring of *c*-subunits (Angevine et al., 2003; Steed and Fillingame, 2008; Steed and Fillingame, 2009). To date, bioinformatic and experimental studies have indicated that the *a*-subunit does not provide any specificity to the enzyme, relegating the responsibility of ion choice to the *c*-ring (Dimroth et al., 2006; Schlegel et al., 2012; Stock et al., 2000; Weber and Senior, 2003). Extensive work on *c*-rings from both $Na^+$- and $H^+$-driven enzymes suggests that all ATPases select $H^+$ over $Na^+$, and that $H^+$ selectivity is an ancestral feature of the F-ATPases (Krah et al., 2010; Leone et al., 2010; Meier et al., 2009; Pogoryelov et al., 2009).

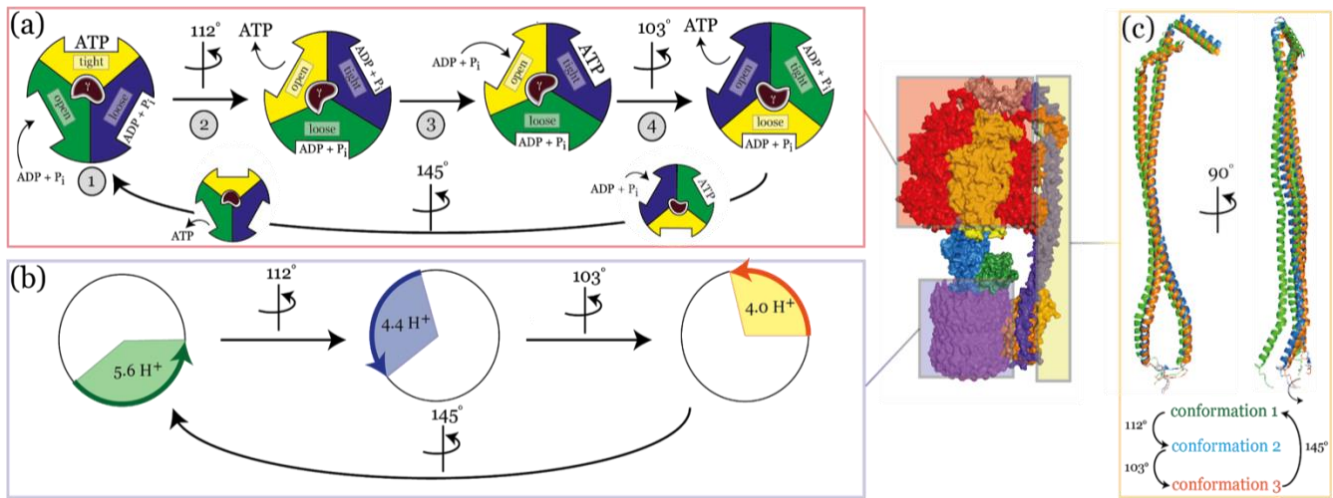

**Figure 3.** Illustration of one full revolution of chloroplast *(a)* F1; *(b)* Fo; and *(c)* the peripheral stalk connection between the two motors. As shown in Figure 2, image processing revealed three primary rotary states in intact chloroplast F-ATPase, separated by 112o, 103o, and 145o (Hahn et al., 2018). *(a)* A full revolution of F1 results in the generation of three ATP molecules, one at each of the three catalytic *β*-subunits via Paul Boyer's proposed 'binding change mechanism' (Boyer 1997; Boyer at al., 1977). The three *β*-subunits cycle between 'open', 'tight', and 'loose' states. Tracking a single *β*-subunit through a full revolution: (1) ADP + Pi arrive and bind at the 'open' site; (2) upon arrival of ADP and phosphate, the 'open' subunit rotates into the 'loose' confirmation, followed by (3) a second rotary power stroke into the 'tight' conformation, in which the catalysis into ATP takes place. (4) The final substep reverts the subunit back to its 'open' conformation, releasing the newly synthesized ATP. *(b)* Rotation angles of the Fo *c*-ring rotor relative to subunit *a*. Primary states of the intact motor do not each correspond to an integral number of *c*-subunits (and consequently, neither to an integral number of ion translocations). *(c)* Two orthogonal views of superposed peripheral stalks, illustrating the flexible coupling between Fo and F1; this flexibility allows for the observed 'symmetry mismatch' between the paired motors (see Figure 4).

This baseline selectivity, however, is neither strong enough to overcome the $Na^+$ excess in extant natural environments nor weak enough to reliably and exclusively couple the enzyme to sodium flow. Therefore, any form of ion selectivity – that is, any non-promiscuous coupling to either *pmf* or *smf* – requires a deviation from the baseline state: to enhance $Na^+$ binding, the binding sites tend towards a higher number of polar amino acids; to favor $H^+$, the sites become more hydrophobic (Krah et al., 2010; Leone et al., 2015; Schulz et al., 2013). Shifts in both directions have been well-characterized across a wide range of bacteria and archaea. For instance, in the alkalophilic bacterium *B. pseudofirmus*, the only hydrophilic residue around the binding sites is the conserved Glu, which is common to all F-ATPases (Pogoryelov et al., 2009; Symersky et al., 2012; Vollmar et al., 2009; Zhou et al., 2015), leading to a strong $H^+$ selectivity (Preiss et al., 2010; Schlegel et al., 2012). Swinging to the other extreme, the well-characterized physiological $Na^+$ coupling in the bacterial species *I. tartaricus* and *E. hirae* arises from three cation-coordinating polar side chains (Meier et al., 2009; Schlegel et al., 2012). A broader overview of the spectrum of ion selectivity across the prokaryotic phylogeny is provided in Figure 4.

Ion flow across biological membranes maintains a wide range of processes in all living cells, from archaea to humans. Of these, one of the first, and most fundamental, is ATP synthesis. Understanding the origin and evolution of bioenergetics therefore relies crucially on the mechanisms adapted by ATP synthases to couple their function to distinct energy sources. Despite the bias towards $H^+$ selectivity in most extant organisms (including all eukaryotes), $Na^+$-dependent ATPases are scattered amongst many branches of the phylogenetic tree. Given the specific identity of the $Na^+$ binding sites across lineages (Mulkidjanian et al., 2009), this provides compelling evidence for the evolutionary primacy of sodium bioenergetics.

A potential implication of an ancestral $Na^+$-dependent ATPase is that the predecessor of modern, proton and sodium-impermeable membranes could have been sodium-impermeable but proton-permeable. One model for early cell evolution is that membranes made from single-chain amphiphiles, such as fatty acids and their derivatives, gradually transitioned to those made from phospholipids (Budin and Szostak, 2011). Fatty acids can be synthesized abiotically and self-assemble into membrane vesicles (Hargreaves and Deamer, 1978) that are capable of spontaneous growth and division (Zhu and Szostak, 2009). Membrane-bound fatty acids switch between deprotonated and protonated forms, with the latter being neutral and capable of rapid (~ms) flip-flop (Kamp et al., 1995) . Through cycles of flip-flop and proton release, fatty acids thus act as protonophores. In contrast, ion permeation in pure phospholipid membranes is mediated by direct ion partitioning or formation of transient pores in the bilayer (Paula et al., 1996).  For this reason, free fatty acids act as natural uncouplers in mitochondria (Di Paola and Lorusso, 2006).  Fatty acids can also transport sodium and other larger cations as salts, but much more slowly than for protons (Chen and Szostak, 2004). Thus, proton-based bioenergetics could have been a challenge for primitive cell membranes with substantial amounts of free fatty acids or high intrinsic leakiness, thus favoring the early utilization of sodium as a less permeable cation. The coupling of transmembrane proton flow to bioenergetic cellular processes would have then arose later, and on several independent occasions, along life's evolutionary path (Mulkidjanian et al., 2008).

| Species | Rotor size | Ion | Reference |
|---|---|---|---|
| *Acetobacterium woodii* | 11 | Na+ | Bogdanović et al., 2019 |
| *Anabaena sp. PCC 7120* | 14 | H+ | Pogoryelov et al., 2007 |
| *Bacillus PS3* | 10 | H+ | Mitome et al., 2004 |
| *Bacillus pseudofirmus* | 13 | H+ | Preiss et al., 2013 |
| *Bacillus sp. TA2.A1* | 13 | H+ | Cook et al., 2003 |
| *Bos taurus*** | 8 | H+ | Giraud et al., 2012 |
| *Burkholderia pseudomallei* | 17 | H+ | Schulz et al., 2017 |
| *Chlamydomonas reinhardtii** | 13 | H+ | zu Tittingdorf et al., 2004 |
| *Clostridium paradoxum* | 11 | H+ | Meier et al., 2006 |
| *Escherichia coli* | 10 | H+ | Jiang et al., 2001 |
| *Fusobacterium nucleatum* | 11 | Na+ | Schulz et al., 2013 |
| *Gloeobacter violaceus* | 15 | H+ | Pogoryelov et al., 2007 |
| *Homo sapiens*** | 10 | H+ | Song et al., 2018 |
| *Ilyobacter tartaricus* | 11 | Na+ | Meier et al., 2005 |
| *Methanopyrus kandleri* | 13 | H+ | Pogoryelov et al., 2007 |
| *Mycobacterium phlei* | 9 | H+ | Preiss et al., 2015 |
| *Pichia angusta*** | 10 | H+ | Vinothkumar et al., 2016 |
| *Paracoccus denitrificans* | 12 | H+ | Morales-Rios et al., 2015 |
| *Polytomella sp. Pringsheim 198.80*** | 10 | H+ | Leone and Faraldo-Gómez, 2016 |
| *Propionigenium modestum* | 11 | Na+ | Meier et al., 2005 |
| *Saccharomyces cerevisiae*** | 10 | H+ | Symersky et al., 2012 |
| *Spinacia oleracea** | 14 | H+ | Hahn et al., 2018 |
| *Spirulina platensis* | 15 | H+ | Pogoryelov et al., 2005 |
| *Synechococcus sp. PCC 6301* | 14 | H+ | Pogoryelov et al., 2007 |
| *Synechococcus sp. PCC 6716* | 14 | H+ | Pogoryelov et al., 2007 |
| *Synechococcus sp. PCC 7942* | 14 | H+ | Pogoryelov et al., 2007 |
| *Synechococcus sp. SAG 89.79* | 13 | H+ | Pogoryelov et al., 2007 |
| *Synechocystis sp. PCC 6803* | 14 | H+ | Pogoryelov et al., 2007 |
| *Yarrowia lipolytica*** | 10 | H+ | Hahn et al., 2016 |

**Table 1.** Experimentally determined rotor ring stoichiometries and ion selectivities in ATP synthases across the tree of life (* denotes chloroplasts; ** denotes mitochondria). Phylogenetic relationships between organisms listed are provided in Figure 4.

Despite much progress in our understanding of membrane mechanics and bioenergetics in recent years, the origin and evolution of biomembranes and the proteins that live within them remains an open, into which the reconstruction of ATP synthase evolution can provide insight (Mulkidjanian et al., 2008).

## If the ring fits: variation in rotor size and stoichiometry

During ATP synthesis, the Fo motor converts the energy stored in the transmembrane ion gradient (either the *pmf* or *smf,* as discussed in the previous section) into torque against the catalytic F1 head (Junge et al., 2009; Martin et al., 2018). In reverse, the F1 head uses the chemical energy released from ATP hydrolysis into torque against Fo, converting its partner into an ion pump (Noji et al., 1997). Whichever way the wheels are turning, a revolution of F1 is tightly coupled mechanically to a revolution of Fo via the central stalk. Across species and cell types, the F1 head invariably contains three catalytic *b*-subunits, and accordingly completes its revolution in three 120° steps, or 'power strokes' (Martin et al., 2018). The intermediate rotary states, then, are determined by the stoichiometry of Fo's rotor ring, that is, the number of *c*-subunits (*n*) that make up the ring.

This nearly always leads to a symmetry mismatch, as *n* is rarely a multiple of 3; one of the rare exceptions to this seems to be the c15 ring of the alkalophilic bacterium *S. platensis* (Pogoryelov et al., 2005). This, as might be expected, leads to rotary states that are not separated by exactly 120° (Hahn et al., 2018). Instead, the three steps are postulated to represent local energy minima in the catalytic cycle that take into account the whole F1Fo complex; as such, they also often do not correspond to integral numbers of *c*-subunits. For example, experiments on the autoinhibited chloroplast ATP synthase showed three primary rotary substates separated by 103°, 112°, and 145°, corresponding to 4, 4.4, and 5.6 *c*-subunits, respectively (Hahn et al., 2018).

Each *c*-subunit contains one binding site for a membrane-crossing ion, and translocates one ion per revolution – or, per ATP synthesized or hydrolyzed. Due to the tight coupling between the two motors, the stoichiometry of the rotor ring, the number of *c*-subunits, theoretically determines the ratio of ions per ATP ($n/3$). The free energy of ATP hydrolysis DG is approximately -50 kJ/mol; taking into account the energetic environment in the spinach chloroplast, transitioning between the three primary substates listed above corresponds to free energy changes of -43.7, -48.1, and -61.2 kJ/mol (Hahn et al., 2018, Kühlbrandt., 2019). The primary substates in *E. coli* ATP synthase were shown to hold to a similar principle (Hahn et al., 2018; Yanagisawa and Frasch, 2017). In mitochondria, the ion-to-ATP coupling ratio is related to the 'P/O ratio', the number of ATP synthesized for each pair of electrons (Ferguson 2010).

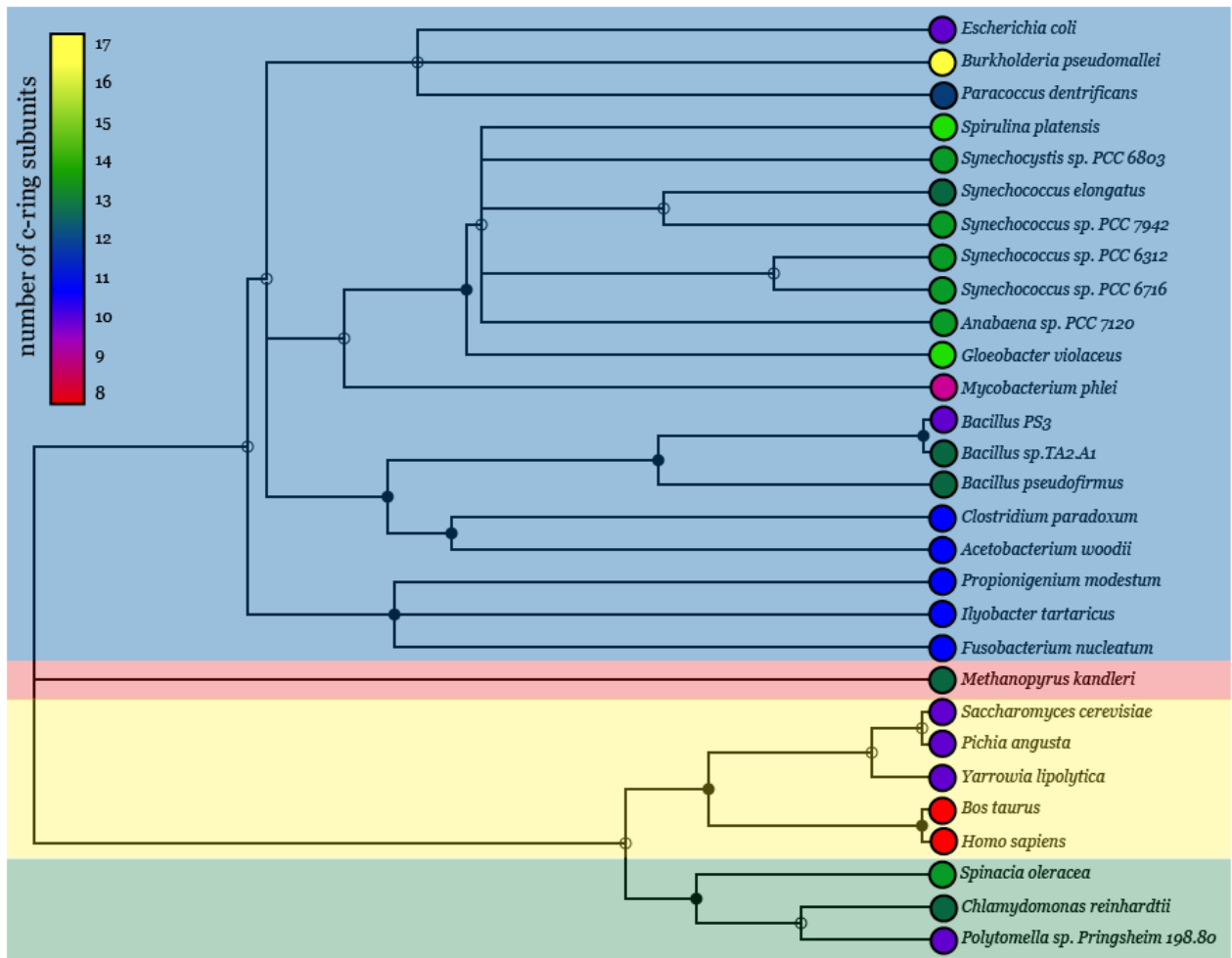

**Figure 4.** Phylogenetic distribution of *c*-ring stoichiometries. Rooted phylogeny of organisms with experimentally determined *c*-ring structures obtained through the TimeTree knowledge base using the method outlined in Hedges et al., 2015. Leaves are marked with colored circles according to the number of subunits found in the ATP synthase *c*-ring. Shading corresponds to whether stoichiometries were determined for bacteria (blue), archaea (red), or eukaryotic mitochondria (yellow) or chloroplasts (green). Stoichiometries and corresponding references are also provided in Table 1.

Rotor ring stoichiometry ranges from 8 *c*-ring subunits in mammalian mitochondria (Zhou et al., 2015; Watt et al., 2010) to 17 in the human pathogen *Burkholderia pseudomallei* (Schulz et al., 2017). In bacteria, this stoichiometry was shown to be independent of growth conditions (Pogoryelov et al., 2012). Based on an analysis of the primary structure of ATP synthase *c*-subunits, ring size was shown to depend on the sequence of amino acids in the contact region between adjacent subunits; this interaction motif is a highly conserved set of glycine (or sometimes alanine) repeats 'GxGxGxGxG' in the N-terminal α-helix (Mackenzie et al., 1997; Preiss et al., 2013; Preiss et al., 2014). This sequence is genetically programmed, which explains why *c*-ring stoichiometry is fixed for members of a given species, even as it varies across different species (see Table 1 and Figure 4).

The number of $c$-subunits in the rotor ring seems to be an adaptation to the physiological environment the ATPase resides in. For instance, the large $c$-ring in *B. pseudomallei* has a high ion-to-ATP ratio, making the enzyme an efficient ion-pump and allowing the pathogen to deal with the hostile acidic host phagosome (Schulz et al., 2017). The overwhelming majority of studies have demonstrated that $c$-ring stoichiometries within a given species do not vary with external conditions like medium pH, host membrane composition, or carbon source (Davis and Kramer, 2020; Fritz et al., 2008; Meier et al., 2005; Meier and Dimroth, 2007; Meier et al., 2007; zu Tittingdorf et al., 2004; but see also, e.g., Schemidt et al., 1998). However, genetically engineered rotors of modified size have been shown to function in both 'directions' within their host cell -- that is, edited F1Fo complexes can both synthesize and hydrolyze ATP. Further, spectroscopic evidence in *I. tartaricus* (wildtype $n$ = 11) implies that the central stalk can interact with rotor rings comprised of a range of $c$-subunits (11 < $n$ < 15) with affinity comparable to that of the wild-type (Pogoryelov et al., 2012). Biochemical and structural evidence also suggest that single mutants in the $c$-subunit of F-ATPases can alter rotor ring stoichiometry, studies suggest that evolutionary adaptation in rotor ring size is largely independent of the mechanical coupling of Fo to the F1 head (Pogoryelov et al., 2012).

Interestingly, changes in rotor ring size are also hypothesized to have occurred as an adaptation to selective pressure on motor torque in the bacterial flagellar motor (BFM), the only other known ion-driven rotary nanomotor. Structural modifications of the BFM rotor, however, involve far more complex genetic maneuvers than in F-ATPases (Beeby et al., 2016; Chaban et al., 2018); as such, bacteria instead modify their *stator* stoichiometry in response to more immediate challenges (Lele et al., 2013; Nirody et al., 2019; Nord et al., 2017; Wadhwa et al., 2019). However, the relative ease of F-ATPases rotor ring modification suggests that a series of gradual changes in $c$-ring stoichiometry may be a feasible 'strategy' to adjust ion-to-ATP ratios in novel environments within a relatively short evolutionary timescale (e.g., in host-pathogen relationships).

## It takes two: dimerization in mitochondrial ATPases

Unlike those found in prokaryotic and chloroplasts (but see, e.g., Daum et al., 2006; Rexroth et al., 2004), mitochondrial F-ATPases (mtF1Fo) all form dimers in the membrane. However, in accordance with the large variation found within the eukaryotic lineages, mitochondrial ATPases vary quite widely with respect to their subunit composition. Consequently, different dimerization strategies are utilized across eukaryotic taxa, with four classes of dimers having been characterized thus far: Type I, found in animals and fungi (Davies et al., 2011; Davies et al., 2012; Dudkina et al., 2006; Strauss et al., 2008); Type II, found in most unicellular green algae except the *Euglena*, which contain Type IV dimers (Blum et al., 2019; Cano-Estrada et al., 2010; Dudkina et al., 2010; van Lis et al., 2005; van Lis et al., 2007; Vazquez-Acevedo et al., 2006); Type III, found in the ciliates (Allen, 1995; Mühleip et al., 2016); and Type IV, found in trypanosomes and *Euglena* (Mühleip et al., 2017). We note that, as new research is constantly plumbing new depths with respect to diversity within the unicellular eukaryotes, the discovery and classification of many more types of F-ATPase dimers is likely. There exists no apparent homology between the four classes, and the major distinction between them is the subunit composition of the dimerization interface. Variations in this interface can cause large changes in the angle observed between central stalks of the two monomeric partners, ranging from close to 0° (Type III dimers; Mühleip et al., 2016) to approximately 90° (Type I dimers; Davies et al., 2011).

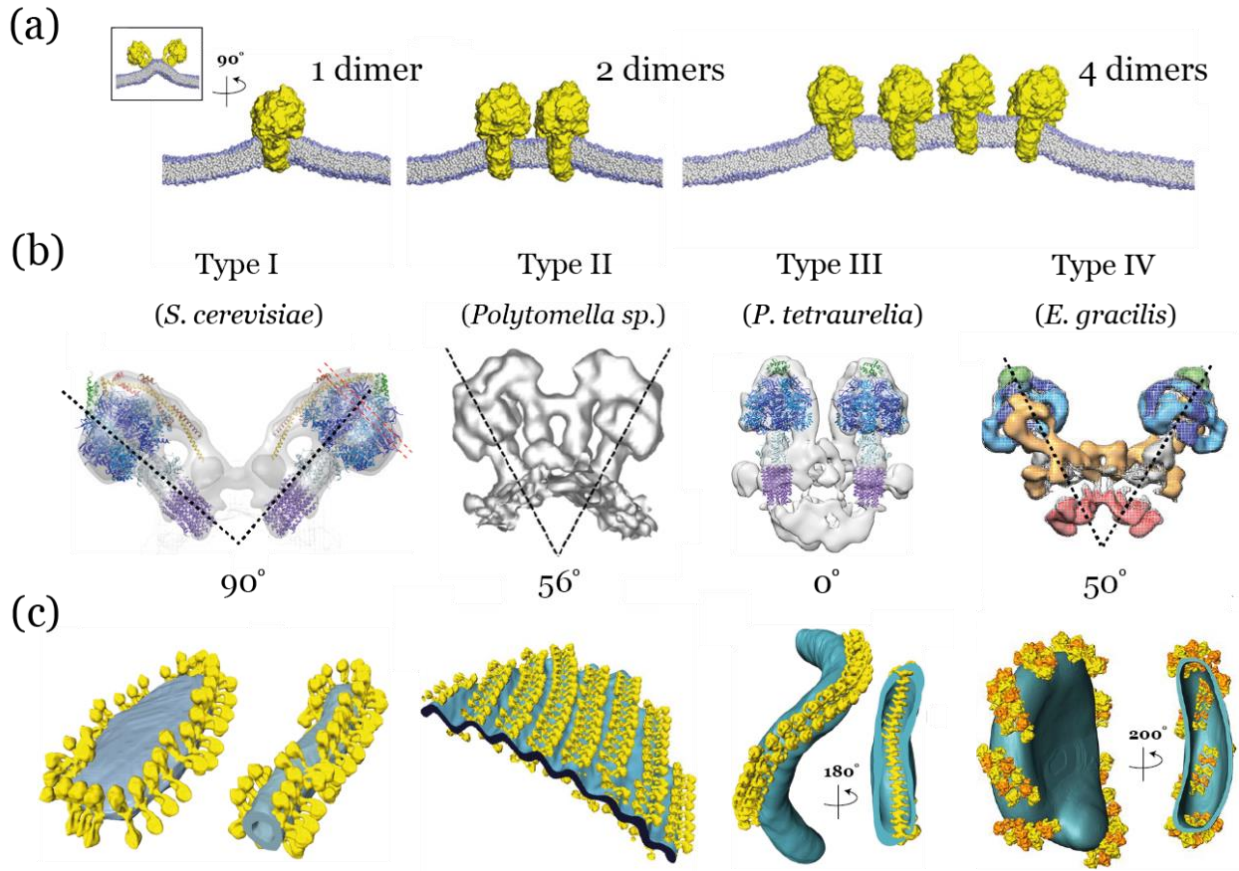

**Figure 5.** Structure of mitochondrial ATP synthase dimers. *(a)* Formation of dimer rows relieves strain caused by dimerization-induced local membrane curvature. Perspective slices through simulated membrane patches illustrate how curvature profile changes when one, two, or four ATP synthase dimers assemble into a row (Davies et al., 2012). *(b)* Subtomogram images of ATP synthase dimers and *(c)* surface representations of cristae structure from representative species of each dimerization class: Type I, *Saccharomyces cerevisiae* (Davies et al., 2012); Type II, *Polytomella sp.* (Blum et al., 2019); Type III, *Paramecium tetraurelia* (Mühleip et al., 2016); Type IV, *Euglena gracilis* (Mühleip et al., 2017).

The dimerization of ATP synthase is tied to the formation of the characteristic folds, or cristae, of the inner mitochondrial membrane (IMM). The larger-scale organization of mtF1Fo dimers within the IMM is determined by the nature of the dimerization interface. Cryo-electron tomograms in fungi and metazoans have revealed that Type I dimers are organized into long (~1mm) rows that align themselves with the sharply curved cristae edges (Davies et al., 2012; Strauss et al., 2008). Mutation studies in yeast have concretely connected the proper development of cristae to F-ATPase dimerization (Davies et al., 2012; Paumard et al., 2002). Row- or ribbon-like formations seem to be a universal feature in mitochondrial F-ATPases. In addition to the aforementioned Type I dimer rows, Type II and Type IV dimers, which exhibit an 'intermediate' dimerization angle of ~55°, form closely spaced helical ribbons that wrap around the edges of the disk-shaped cristae of the unicellular green algae and trypanosomes (Blum et al., 2019; Mühleip et al., 2016). Molecular simulation studies have suggested that dimers spontaneously form rows in order to minimize the overall elastic strain on the IMM which arises from the local curvature imposed upon the membrane by ATPase dimerization (Anselmi et al., 2016; Davies et al., 2012). This mechanism is independent of lipid composition or any particular protein-protein interaction.

The 'U-shaped' Type III dimers found in ciliates, however, form a dimer angle close to 0° -- that is, the ATPase monomers are parallel to each other. This conformation does not as obviously lend itself to the membrane deformation hypothesis as the angles of their more bent cousins. Despite this, ciliate mitochondrial membranes also show tight helical rows of dimerized ATPases, which give rise to tube-like cristae (Mühleip et al., 2016). An analogous model for the formation of these structures was proposed in which the curvature of the IMM is not simply a consequence of dimer formation, but a more specific interaction between the U-shaped dimers during row assembly. In particular, the peripheral stalks of the dimers bend the IMM, driving row formation to relieve local bending energy as in mitochondria containing V-shaped dimers (Mühleip et al., 2016). An overview of the characterized classes of ATPase dimers and their respective cristae morphologies are shown in Figure 5.

Studies have also illustrated the importance of cristae structure on mitochondrial, and consequently on cellular and organismal, health (Bornhövd et al., 2006; Davies et al., 2012; Paumard et al. 2002). For instance, the breakdown of cristae morphology in fungal mitochondria has been associated with senescence, and comparison of cryo-ET images from young and aging cultures of the fungal mold *Podospora anserina* have linked the loss of normal lamellar cristae structure to the breakdown of Type I dimer rows (Daum et al., 2013). Studies in metazoan aging models demonstrated clear functional links between aging-related degradation and dimer-induced cristae ultrastructure (Brandt et al., 2017). Furthermore, in mice, changes in membrane structure were shown to be organ-dependent, such that different organs exhibit different degrees of resilience against aging. For example, stacking of cristae in cardiac cells is packed more tightly than in other tissues, such as in the liver or kidney (Brandt et al., 2017).

One seemingly obvious effect of cristae packing structure is the increase of surface area in the inner mitochondrial membrane, which is important for increasing the flux of molecules across the membrane. However, elimination of regular cristae structure in yeast via mutation of subunits at the dimerization interface did not noticeably reduce inner membrane surface area, though mutants showed a significantly lower growth rate (Davies et al., 2012; Paumard et al., 2002). Other studies have implicated cristae in maintaining the mitochondrial membrane potential (Bornhövd et al., 2006; but see also Saddar et al., 2008), suggesting that cristae serve as proton-trapping 'microcompartments' to aid the conveniently arranged rows of nearby ATP synthases (Daum et al., 2013; Davies et al., 2012; Davies et al., 2018; Kühlbrandt, 2019). In *Drosophila*, dimerization of ATP synthase has been shown to be essential both for stem cell differentiation during development as well as to ward off aging-related degradation (Brandt et al., 2017; Teixeira et al., 2015). The mechanisms behind these effects are not yet known, and more investigation is needed to clarify the link between mtF1Fo dimer formation and age-related changes – for instance, whether dimer dissociation is a causal agent, effect, or byproduct of senescence.

## Dancing in the bilayer: interaction of ATPases with membrane lipids

Like all membrane-embedded proteins, ATPases live in complex surroundings whose lipid building blocks of which have co-evolved with their resident proteins. Thus, several open questions in bioenergetics research center around the effects and drivers of the local membrane environment (Figure 6). In eukaryotes, distinct membrane compartments (van Meer et al, 2008) and even bilayer leaflets (Lorent et al., 2020) have their own characteristic properties based on the lipid biosynthetic and transport processes that determine their composition. The IMM is particularly unique: it is almost completely bereft of sterols and saturated sphingolipids, molecules that order and provide rigidity to the membranes and other cellular compartments. The most abundant lipid of the IMM inner leaflet is phosphatidylethanolamine (PE), while phosphatidylcholine is dominant in the outer leaflet, as it is in the rest of the cell (Colbeau et al., 1971). Even more striking is the abundance of cardiolipin (CL), a tetracyl anionic phospholipid that comprises > 20% of IMM membranes and is exclusive to the mitochondria. Like PE, CL is enriched to the inner leaflet of the IMM (Krebs et al., 979), where it comprises an even larger fraction of the lipids.

The composition of the IMM harkens back to its bacterial origins. In proteobacteria, PE is the dominant phospholipid while CL is also synthesized, but at lower levels -- CL comprises approximately 5% of all lipids in *E. coli*, where it is non-essential (Nishijima et al., 1988). Thus eukaryotic adaptations in ATPase structure and mitochondrial morphology likely co-evolved alongside large-scale changes in its IMM composition, particularly the increase in CL content. Notably, the radius of curvature that characterizes cristae structures (~15 nm) is much smaller than that of bacterial poles (~0.5 μm), a geometric constraint that potentially underlies adaptations in both ATPase organization and lipid composition.

The biophysical properties of CL are unique among lipids and have likely led to its close association with ATPase dimers in the IMM. Containing four voluminous lipid chains, CL is regarded as a non-bilayer lipid as it cannot form planar membranes by itself due to its conic shape. Pure CL instead forms hexagonal cubic and other non-lamellar phases (Ortiz et al., 1999). When incorporated into membranes, this molecular morphology imposes an intrinsic curvature, thereby facilitating the bending of highly curved structures (Alimohamadi and Rangamani, 2018; Chabanon, et al., 2017). Many of the insights for this function came from studies in bacteria, where imaging with the dye Nonyl Acridine Orange has revealed the heterogeneous distribution on CL on cell poles and division septa (Mileykovskaya and Dowhan, 2000; Kawai et al., 2004; Oliver et al., 2014). More recent studies utilizing model membranes have demonstrated the intrinsic capability of CL to localize to thin membrane tubules (Beltrán-Heredia et al., 2019). Because the diameter of these tubules is much smaller than that of bacterial cell, it has been proposed that domains of CL must 'sense' biologically-relevant curvatures (Huang et al., 2006), as is the model for curvature sensing proteins (Ramamurthi et al., 2009). In the IMM, however, curvature is much higher at cristae edges where ATPase dimers reside, so any higher order organization is not necessary for sorting of CL to these sites. Therefore, it is generally accepted that CL is enriched at these sites, and it is possible that its enrichment allowed for the evolution of cristae structures. In one striking demonstration, the ability of mutant *E. coli* to form cristae-like internal membrane upon membrane protein overexpression was shown to be dependent on the presence of CL (Carranza et al., 2017).

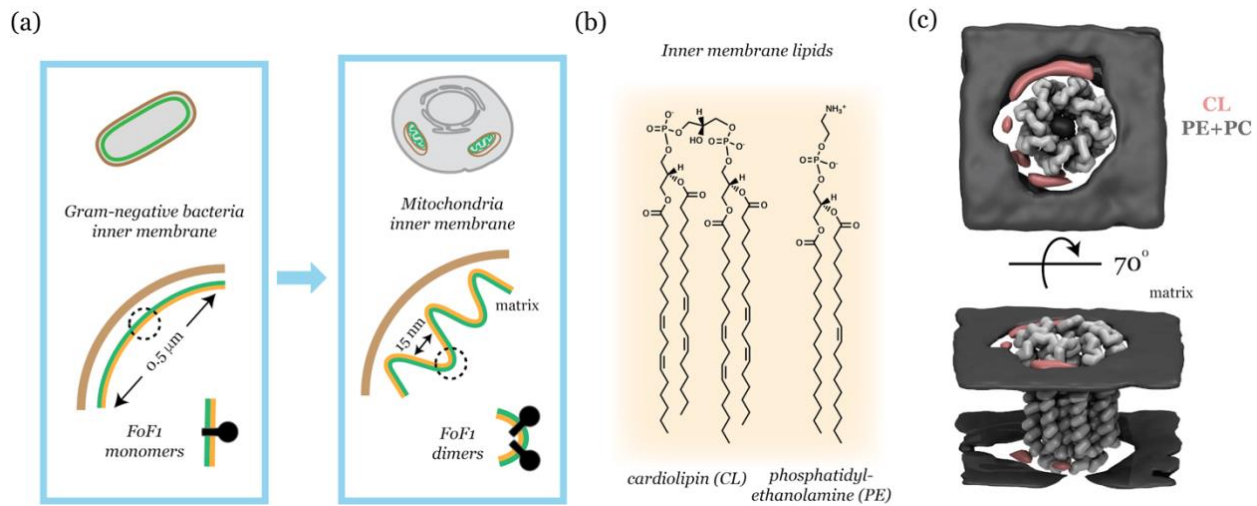

**Figure 6.** Lipids associated with ATP synthase. *(a)* The evolution of membrane topology from the lightly-curved inner membrane of gram-negative proteobacteria to the highly-curved inner membrane of eukaryotic mitochondria. For both, the inner membrane contains asymmetric distribution of lipids that provide spontaneous membrane curvature. In mitochondria, asymmetry is more dramatic, and likely acts alongside the dimerization of ATP synthase to support cristae structure. *(b)* Characteristic lipids of the inner mitochondrial membrane include CL and PE. CL is shown with four linoleic acid chains, as is typical in many mammalian tissues. These lipids are also enriched on the inner leaflet of the membrane, as they are gram-negative bacteria (Bogdanov et al., 2020). *(c)* Coarse-grained MD simulations of the bovine $c_8$-ring in a model inner membrane suggest transient interactions with CL, whose mean density distribution is shown in pink. It has been hypothesized that these interactions promote the rotation of the ring during enzyme cycles. Figure adapted from (Duncan, et al., 2016).

Another potentially conserved function for CL is to recruit proteins to highly curved membranes, potentially through electrostatic interactions. In bacteria, deletion of cardiolipin synthase leads to the mislocalization of some polar-localized membrane proteins (Romantsov et al., 2010), suggesting a mechanism by which CL-protein interactions serve as a localization cue. In mitochondria, CL interactions have likely evolved as a mechanism for localizing mtF1Fo to cristae edges. Before models of cristae organization were developed, biochemical studies of isolated mitochondria from bovine found that mtF1Fo ATPase remained associated with 2.5 mol of CL, higher than all other lipids combined (Eble et al., 1990). More recent work has used genetic disruption of CL levels in several organisms to demonstrate a broad role for cristae organization (Sakamoto et al., 2012). In *Drosophila*, isolated mitochondria from flight muscle of mutants with reduced amounts of CL show slightly enlarged cristae with far fewer ATPases localizing to their edges (Acehan et al., 2011). This suggests that, at least in flies, CL has a major role in anchoring ATPases to cristae edges, but it is still unclear if CL drives cristae curvature or it acts through recruiting ATPase dimers. Better mechanical modeling of cristae membranes is needed to further understand the relative contributions both lipid composition and ATPase structure have on the IMM.

In addition to its association with curved membranes, specific lipids could act as modulators of ATPase function through biophysical effects on its mechanical cycle. Changes in CL levels modulate the respiratory cycle in biochemically isolated mitochondrial membranes (Koshkin and Greenberg, 2002), suggesting a role in dictating ETC enzyme functions. An early hypothesis is that the anionic CL can act as an intermediate proton-trap that facilitates proton translocation across the bilayer (Haines and Dencher, 2002), which would also be based on transient interactions with the F-ATPase rotor ring. Unlike other ETC complexes (Paradies et al., 2014), CL has not been observed in ATPase structures, so where would this association take place? The *c*-ring of metazoan ATPases feature a conserved trimethylated lysine residue (Walpole et al., 2015) that, in simulations, binds transiently to CL molecules (Duncan et al., 2016). This model (Figure 6c) suggests that CL interactions may have an additional stabilizing effect for the *c*-ring, which completes 100 rotations in the bilayer every second.

One possibility is that direct association with lipids can reduce the friction of c-ring rotation, although the mechanics of those proposals have not been fully explored. More generally, lipids mediate the bulk hydrodynamic properties of the membranes that ATPases reside in, so aspects of lipid composition that affect membrane viscosity would affect rotational friction (Saffman and Delbruck, 1975). In metazoans, CL acyl chains are composed of highly-fluidizing polyunsaturated fatty acids (Pennington et al., 2019), which are incorporated via the acyl chain remodeler Tafazzin. Mutations in the gene encoding Tafazzin in humans, is the cause of Barth syndrome and other metabolic diseases. Unsaturation of phosphatidyl lipids (Budin et al., 2018) and PE head group composition (Dawaliby et al., 2016) can also impact membrane viscosity and could similarly influence the rotational speeds of ATPase.

## Rolling forward: conclusions and future work

Energetic considerations inform the structure, function, and evolution of cells and organisms, making the ion-translocating ATP synthase an essential and ubiquitous piece of molecular machinery. As such, small changes or defects can have catastrophic consequences, particularly in cell types whose functions necessitate the production of copious amounts of energy. These include, predictably, neurons and muscle. Indeed, several neuromuscular and neurodegenerative diseases have been linked to mitochondrial disorder, and, in particular, to ATP synthase malfunction (see e.g., Dautant et al., 2018 and Kühlbrandt, 2019 for a more comprehensive discussion). Several recent high-resolution insights into the structure of ATP synthase suggest avenues towards the development of therapies for such disorders. Further, investigations of the similarities and divergences between human mitochondrial ATPases and those of our parasites and pathogens point towards novel (and, importantly, strongly conserved) targets for antibiotic development, which may prove particularly salient in combating multidrug resistant organisms (Mühleip et al., 2016; Mühleip et al., 2017; Preiss et al., 2015).

The components of ATP synthase, especially its catalytic domains, are widely conserved from prokaryotes to plants and metazoans. The F-type ATPases in particular are responsible for bioenergetic processes in all domains of life, with members found in archaea, bacteria, and eukaryotes (Koumandou and Kossida, 2014; Mulkidjanian et al., 2007). The conserved structure and mechanism of these enzymes strongly suggest that they predate the divergence of these lineages (Rand et al., 2004), likely making ATP synthase, alongside the ribosome, among the most ancient molecular motors (Sousa et al., 2016). Though developing and testing models of early evolution is inherently challenging, there is much to be gained from understanding the origin and evolution of ATP synthase. The spread of this enzyme across the tree of life implies that the reconstruction of this lineage may also provide valuable insight into the origins of semi-permeable membranes, as these likely occurred as a co-evolutionary process (Mulkidjanian et al., 2008). ATP synthase does not function in a vacuum -- its interactions with its lipid environment and proteinaceous neighbors has shaped its structure and function. To this end, we hope that this review highlights the importance of considering how the complex environment of membrane proteins has impacted their function and evolutionary trajectory.

## References

Acehan, D., Malhotra, A., Xu, Y., Ren, M., Stokes, D. L., & Schlame, M. (2011). Cardiolipin affects the supramolecular organization of ATP synthase in mitochondria. *Biophysical Journal*, 100(9), 2184-2192.

Alimohamadi, H., & Rangamani, P. (2018). Modeling membrane curvature generation due to membrane–protein interactions. *Biomolecules*, 8(4), 120.

Allen, R.D (1995). Membrane tubulation and proton pumps. *Protoplasma*, *189*(1-2), 1-8.

Angevine, C. M., Herold, K. A., & Fillingame, R. H. (2003). Aqueous access pathways in subunit a of rotary ATP synthase extend to both sides of the membrane. *Proceedings of the National Academy of Sciences*, *100*(23), 13179-13183.

Anselmi, C., Davies, K. M., & Faraldo-Gómez, J. D. (2018). Mitochondrial ATP synthase dimers spontaneously associate due to a long-range membrane-induced force. *Journal of General Physiology*, *150*(5), 763-770.

Banfield, J. F., & Marshall, C. R. (2000). Genomics and the geosciences. *Science*, *287* (5453), 605-606.

Bason, J. V., Montgomery, M. G., Leslie, A. G., & Walker, J. E. (2014). Pathway of binding of the intrinsically disordered mitochondrial inhibitor protein to F1-ATPase. *Proceedings of the National Academy of Sciences*, *111*(31), 11305-11310.

Battistuzzi, F. U., Feijao, A., & Hedges, S. B. (2004). A genomic timescale of prokaryote evolution: insights into the origin of methanogenesis, phototrophy, and the colonization of land. *BMC Evolutionary Biology*, *4*(1), 44.

Beeby, M., Ribardo, D. A., Brennan, C. A., Ruby, E. G., Jensen, G. J., & Hendrixson, D. R. (2016). Diverse high-torque bacterial flagellar motors assemble wider stator rings using a conserved protein scaffold. *Proceedings of the National Academy of Sciences*, *113*(13), E1917-E1926.

Beltrán-Heredia, E., Tsai, F. C., Salinas-Almaguer, S., Cao, F. J., Bassereau, P., & Monroy, F. (2019). Membrane curvature induces cardiolipin sorting. *Communications Biology*, 2(1), 1-7.

Blum, T. B., Hahn, A., Meier, T., Davies, K. M., & Kühlbrandt, W. (2019). Dimers of mitochondrial ATP synthase induce membrane curvature and self-assemble into rows. *Proceedings of the National Academy of Sciences*, *116*(10), 4250-4255.

Bogdanov, M., Pyrshev, K., Yesylevskyy, S., Ryabichko, S., Boiko, V., Ivanchenko, P., ... & Dowhan, W. (2020). Phospholipid distribution in the cytoplasmic membrane of Gram-negative bacteria is highly asymmetric, dynamic, and cell shape-dependent. *Science Advances*, 6(23), eaaz6333.

Bogdanović, N., Trifunović, D., Sielaff, H., Westphal, L., Bhushan, S., Müller, V., & Grüber, G. (2019). The structural features of Acetobacterium woodii F-ATP synthase reveal the importance of the unique subunit γ-loop in Na+ translocation and ATP synthesis. *The FEBS Journal*, 286(10), 1894-1907.

Bornhövd, C., Vogel, F., Neupert, W., & Reichert, A. S. (2006). Mitochondrial membrane potential is dependent on the oligomeric state of F1F0-ATP synthase supracomplexes. *Journal of Biological Chemistry*, *281*(20), 13990-13998.

Boucher, Y., Douady, C. J., Papke, R. T., Walsh, D. A., Boudreau, M. E. R., Nesbø, C. L., ... & Doolittle, W. F. (2003). Lateral gene transfer and the origins of prokaryotic groups. *Annual Review of Genetics*, *37*(1), 283-328.

Boyer, P. D. (1997). The ATP synthase—a splendid molecular machine. *Annual review of Biochemistry*, *66*(1), 717-749.

Boyer, P. D. (1993). The binding change mechanism for ATP synthase—some probabilities and possibilities. *Biochimica et Biophysica Acta (BBA)-Bioenergetics*, *1140*(3), 215-250.

Boyer, P. D., Chance, B., Ernster, L., Mitchell, P., Racker, E., & Slater, E. C. (1977). Oxidative phosphorylation and photophosphorylation. *Annual Review of Biochemistry*, *46*(1), 955-966.

Brandt, T., Mourier, A., Tain, L. S., Partridge, L., Larsson, N. G., & Kühlbrandt, W. (2017). Changes of mitochondrial ultrastructure and function during ageing in mice and Drosophila. *eLife*, *6*, e24662.

Budin, I., de Rond, T., Chen, Y., Chan, L. J. G., Petzold, C. J., & Keasling, J. D. (2018). Viscous control of cellular respiration by membrane lipid composition. *Science*, *362*(6419), 1186-1189.

Budin, I., & Szostak, J. W. (2011). Physical effects underlying the transition from primitive to modern cell membranes. *Proceedings of the National Academy of Sciences*, 108(13), 5249-5254.

Carranza, G., Angius, F., Ilioaia, O., Solgadi, A., Miroux, B., & Arechaga, I. (2017). Cardiolipin plays an essential role in the formation of intracellular membranes in Escherichia coli. *Biochimica et Biophysica Acta (BBA)-Biomembranes*, *1859*(6), 1124-1132.

Castresana, J. (2001). Comparative genomics and bioenergetics. *Biochimica et Biophysica Acta (BBA)-Bioenergetics*, *1506*(3), 147-162.

Chaban, B., Coleman, I., & Beeby, M. (2018). Evolution of higher torque in Campylobacter-type bacterial flagellar motors. *Scientific Reports*, *8*(1), 1-11.

Chabanon, M., Stachowiak, J. C., & Rangamani, P. (2017). Systems biology of cellular membranes: a convergence with biophysics. *Wiley Interdisciplinary Reviews: Systems Biology and Medicine*, 9(5), e1386.

Chen, I. A., & Szostak, J. W. (2004). A kinetic study of the growth of fatty acid vesicles. *Biophysical Journal*, 87(2), 988-998.

Cherepanov, D. A., Mulkidjanian, A. Y., & Junge, W. (1999). Transient accumulation of elastic energy in proton translocating ATP synthase. *FEBS Letters*, *449*(1), 1-6.

Colbeau, A., Nachbaur, J., & Vignais, P. M. (1971). Enzymac characterization and lipid composition of rat liver subcellular membranes. *Biochimica et Biophysica Acta (BBA)-Biomembranes*, *249*(2), 462-492.

Cook, G. M., Keis, S., Morgan, H. W., von Ballmoos, C., Matthey, U., Kaim, G., & Dimroth, P. (2003). Purification and biochemical characterization of the F1Fo-ATP synthase from thermoalkaliphilic Bacillus sp. strain TA2. A1. *Journal of Bacteriology*, 185(15), 4442-4449.

Daum, B., Nicastro, D., Austin, J., McIntosh, J. R., & Kühlbrandt, W. (2010). Arrangement of photosystem II and ATP synthase in chloroplast membranes of spinach and pea. *The Plant Cell*, *22*(4), 1299-1312.

Daum, B., Walter, A., Horst, A., Osiewacz, H. D., & Kühlbrandt, W. (2013). Age-dependent dissociation of ATP synthase dimers and loss of inner-membrane cristae in mitochondria. *Proceedings of the National Academy of Sciences*, *110*(38), 15301-15306.

Dautant, A., Meier, T., Hahn, A., Tribouillard-Tanvier, D., Di Rago, J. P., & Kucharczyk, R. (2018). ATP synthase diseases of mitochondrial genetic origin. *Frontiers in Physiology*, 9, 329.

Davies, K. M., Anselmi, C., Wittig, I., Faraldo-Gómez, J. D., & Kühlbrandt, W. (2012). Structure of the yeast F1Fo-ATP synthase dimer and its role in shaping the mitochondrial cristae. *Proceedings of the National Academy of Sciences*, *109*(34), 13602-13607.

Davies, K. M., Blum, T. B., & Kühlbrandt, W. (2018). Conserved in situ arrangement of complex I and III2 in mitochondrial respiratory chain supercomplexes of mammals, yeast, and plants. *Proceedings of the National Academy of Sciences*, *115*(12), 3024-3029.

Davies, K. M., Strauss, M., Daum, B., Kief, J. H., Osiewacz, H. D., Rycovska, A., ... & Kühlbrandt, W. (2011). Macromolecular organization of ATP synthase and complex I in whole mitochondria. *Proceedings of the National Academy of Sciences*, *108*(34), 14121-14126.

Davis, G. A., & Kramer, D. M. (2020). Optimization of ATP Synthase c-Rings for Oxygenic Photosynthesis. *Frontiers in Plant Science*, 10, 1778.

Dawaliby, R., Trubbia, C., Delporte, C., Noyon, C., Ruysschaert, J. M., Van Antwerpen, P., & Govaerts, C. (2016). Phosphatidylethanolamine is a key regulator of membrane fluidity in eukaryotic cells. *Journal of Biological Chemistry*, *291*(7), 3658-3667.

Dibrova, D. V., Galperin, M. Y., & Mulkidjanian, A. Y. (2010). Characterization of the N-ATPase, a distinct, laterally transferred Na+-translocating form of the bacterial F-type membrane ATPase. *Bioinformatics*, *26*(12), 1473-1476.

Dimroth, P. (1997). Primary sodium ion translocating enzymes. *Biochimica et Biophysica Acta (BBA)-Bioenergetics*, *1318*(1-2), 11-51.

Dimroth, P., von Ballmoos, C., & Meier, T. (2006). Catalytic and mechanical cycles in F-ATP synthases. *EMBO Reports*, *7*(3), 276-282.

Dimroth, P., Wang, H., Grabe, M., & Oster, G. (1999). Energy transduction in the sodium F-ATPase of Propionigenium modestum. *Proceedings of the National Academy of Sciences*, *96*(9), 4924-4929.

Di Paola, M., & Lorusso, M. (2006). Interaction of free fatty acids with mitochondria: coupling, uncoupling and permeability transition. *Biochimica et Biophysica Acta (BBA)-Bioenergetics*, 1757(9-10), 1330-1337.

Dudkina, N. V., Oostergetel, G. T., Lewejohann, D., Braun, H. P., & Boekema, E. J. (2010). Row-like organization of ATP synthase in intact mitochondria determined by cryo-electron tomography. *Biochimica et Biophysica Acta (BBA)-Bioenergetics*, *1797*(2), 272-277.

Dudkina, N. V., Sunderhaus, S., Braun, H. P., & Boekema, E. J. (2006). Characterization of dimeric ATP synthase and cristae membrane ultrastructure from Saccharomyces and Polytomella mitochondria. *FEBS Letters*, *580*(14), 3427-3432.

Duncan, A. L., Robinson, A. J., & Walker, J. E. (2016). Cardiolipin binds selectively but transiently to conserved lysine residues in the rotor of metazoan ATP synthases. *Proceedings of the National Academy of Sciences*, *113*(31), 8687-8692.

Eble, K. S., Coleman, W. B., Hantgan, R. R., & Cunningham, C. C. (1990). Tightly associated cardiolipin in the bovine heart mitochondrial ATP synthase as analyzed by 31P nuclear magnetic resonance spectroscopy. *Journal of Biological Chemistry*, *265*(32), 19434-19440.

Elston, T., Wang, H., & Oster, G. (1998). Energy transduction in ATP synthase. *Nature*, *391*(6666), 510-513.

Ethayathulla, A. S., Yousef, M. S., Amin, A., Leblanc, G., Kaback, H. R., & Guan, L. (2014). Structure-based mechanism for Na+/melibiose symport by MelB. *Nature Communications*, *5*(1), 1-11.

Feniouk, B. A., Kozlova, M. A., Knorre, D. A., Cherepanov, D. A., Mulkidjanian, A. Y., & Junge, W. (2004). The proton-driven rotor of ATP synthase: ohmic conductance (10 fS), and absence of voltage gating. *Biophysical Journal*, *86*(6), 4094-4109.

Ferguson, S. J. (2010). ATP synthase: from sequence to ring size to the P/O ratio. *Proceedings of the National Academy of Sciences*, *107*(39), 16755-16756.

Fischer, S., Etzold, C., Turina, P., Deckers-Hebestreit, G., Altendorf, K., & Gräber, P. (1994). ATP synthesis catalyzed by the ATP synthase of Escherichia coli reconstituted into liposomes. *European Journal of Biochemistry*, *225*(1), 167-172.

Forgac, M. (2007). Vacuolar ATPases: rotary proton pumps in physiology and pathophysiology. *Nature Reviews Molecular Cell Biology*, *8*(11), 917-929.

Fritz, M., Klyszejko, A. L., Morgner, N., Vonck, J., Brutschy, B., Muller, D. J., ... & Müller, V. (2008). An intermediate step in the evolution of ATPases–a hybrid F0–V0 rotor in a bacterial Na+ F1F0 ATP synthase. *The FEBS Journal*, *275*(9), 1999-2007.

Gibbons, C., Montgomery, M. G., Leslie, A. G., & Walker, J. E. (2000). The structure of the central stalk in bovine F 1-ATPase at 2.4 Å resolution. *Nature Structural Biology*, *7*(11), 1055-1061.

Giraud, M. F., Paumard, P., Sanchez, C., Brèthes, D., Velours, J., & Dautant, A. (2012). Rotor architecture in the yeast and bovine F1-c-ring complexes of F-ATP synthase. *Journal of Structural Biology*, 177(2), 490-497.

Gledhill, J. R., Montgomery, M. G., Leslie, A. G., & Walker, J. E. (2007). How the regulatory protein, IF1, inhibits F1-ATPase from bovine mitochondria. *Proceedings of the National Academy of Sciences*, *104*(40), 15671-15676.

Grüber, G., Manimekalai, M. S. S., Mayer, F., & Müller, V. (2014). ATP synthases from archaea: the beauty of a molecular motor. *Biochimica et Biophysica Acta (BBA)-Bioenergetics*, *1837*(6), 940-952.

Guo, H., Suzuki, T., & Rubinstein, J. L. (2019). Structure of a bacterial ATP synthase. *eLife*, *8*, e43128.

Hahn, A., Parey, K., Bublitz, M., Mills, D. J., Zickermann, V., Vonck, J., ... & Meier, T. (2016). Structure of a complete ATP synthase dimer reveals the molecular basis of inner mitochondrial membrane morphology. *Molecular Cell*, 63(3), 445-456.

Hahn, A., Vonck, J., Mills, D. J., Meier, T., & Kühlbrandt, W. (2018). Structure, mechanism, and regulation of the chloroplast ATP synthase. *Science*, *360*(6389), eaat4318.

Haines, T. H., & Dencher, N. A. (2002). Cardiolipin: a proton trap for oxidative phosphorylation. *FEBS Letters*, *528*(1-3), 35-39.

Hargreaves, W. R., & Deamer, D. W. (1978). Liposomes from ionic, single-chain amphiphiles. *Biochemistry*, 17(18), 3759-3768.

Hedges, S. B., Marin, J., Suleski, M., Paymer, M., & Kumar, S. (2015). Tree of life reveals clock-like speciation and diversification. *Molecular Biology and Evolution*, 32(4), 835-845.

Hilario, E., & Gogarten, J. P. (1998). The prokaryote-to-eukaryote transition reflected in the evolution of the V/F/A-ATPase catalytic and proteolipid subunits. *Journal of Molecular Evolution*, *46*(6), 703-715.

Hohmann-Marriott, M. F., & Blankenship, R. E. (2011). Evolution of photosynthesis. *Annual Review of Plant Biology*, *62*, 515-548.

Huang, K. C., Mukhopadhyay, R., & Wingreen, N. S. (2006). A curvature-mediated mechanism for localization of lipids to bacterial poles. *PLoS Computational Biology*, *2*(11).

Ihara, K., Abe, T., Sugimura, K. I., & Mukohata, Y. (1992). Halobacterial A-ATP synthase in relation to V-ATPase. *Journal of Experimental Biology*, *172*(1), 475-485.

Imamura, H., Nakano, M., Noji, H., Muneyuki, E., Ohkuma, S., Yoshida, M., & Yokoyama, K. (2003). Evidence for rotation of V1-ATPase. *Proceedings of the National Academy of Sciences*, 100(5), 2312-2315.

Iwabe, N., Kuma, K. I., Hasegawa, M., Osawa, S., & Miyata, T. (1989). Evolutionary relationship of archaebacteria, eubacteria, and eukaryotes inferred from phylogenetic trees of duplicated genes. *Proceedings of the National Academy of Sciences*, *86*(23), 9355-9359.

Jiang, W., Hermolin, J., & Fillingame, R. H. (2001). The preferred stoichiometry of c subunits in the rotary motor sector of *Escherichia coli* ATP synthase is 10. *Proceedings of the National Academy of Sciences*, 98(9), 4966-4971.

Junge, W., Lill, H., & Engelbrecht, S. (1997). ATP synthase: an electrochemical transducer with rotatory mechanics. *Trends in Biochemical Sciences*, *22*(11), 420-423.

Junge, W., Sielaff, H., & Engelbrecht, S. (2009). Torque generation and elastic power transmission in the rotary F O F 1-ATPase. *Nature*, *459*(7245), 364.

Kagawa, Y., & Racker, E. (1966). Partial Resolution of the Enzymes Catalyzing Oxidative Phosphorylation VIII. Properties of a factor conferring oligomycin sensitivity on mitochondrial adenosine triphosphatase. *Journal of Biological Chemistry*, 241(10), 2461-2466.

Kamp, F., Zakim, D., Zhang, F., Noy, N., & Hamilton, J. A. (1995). Fatty acid flip-flop in phospholipid bilayers is extremely fast. *Biochemistry*, 34(37), 11928-11937.

Kawai, F., Shoda, M., Harashima, R., Sadaie, Y., Hara, H., & Matsumoto, K. (2004). Cardiolipin domains in *Bacillus subtilis* marburg membranes. *Journal of bacteriology*, *186*(5), 1475-1483.

Knoll, A. H., & Bauld, J. (1989). The evolution of ecological tolerance in prokaryotes. *Earth and Environmental Science Transactions of The Royal Society of Edinburgh*, *80*(3-4), 209-223.

Kohzuma, K., Dal Bosco, C., Kanazawa, A., Kramer, D. M., & Meurer, J. (2013). A Potential Function for the γ2 Subunit (atpC2) of the Chloroplast ATP Synthase. In *Photosynthesis Research for Food, Fuel and the Future* (pp. 193-196). Springer, Berlin, Heidelberg.

Koshkin, V., & Greenberg, M. L. (2002). Cardiolipin prevents rate-dependent uncoupling and provides osmotic stability in yeast mitochondria. *Biochemical Journal*, *364*(1), 317-322.

Krebs, J. J., Hauser, H., & Carafoli, E. (1979). Asymmetric distribution of phospholipids in the inner membrane of beef heart mitochondria. *Journal of Biological Chemistry*, *254*(12), 5308-5316.

Kühlbrandt, W. (2019). Structure and mechanisms of F-type ATP synthases. *Annual Review of Biochemistry*, *88*, 515-549.

Krah, A., Pogoryelov, D., Langer, J. D., Bond, P. J., Meier, T., & Faraldo-Gómez, J. D. (2010). Structural and energetic basis for H+ versus Na+ binding selectivity in ATP synthase Fo rotors. *Biochimica Et Biophysica Acta (BBA)-Bioenergetics*, *1797*(6-7), 763-772.

Koumandou, V. L., & Kossida, S. (2014). Evolution of the F0F1 ATP synthase complex in light of the patchy distribution of different bioenergetic pathways across prokaryotes. *PLoS Computational Biology*, *10*(9).

Laget, P. P., & Smith, J. B. (1979). Inhibitory properties of endogenous subunit ϵ in the Escherichia coli F1 ATPase. *Archives of Biochemistry and Biophysics*, *197*(1), 83-89.

Lapierre, P., Shial, R., & Gogarten, J. P. (2006). Distribution of F-and A/V-type ATPases in *Thermus scotoductus* and other closely related species. *Systematic and applied microbiology*, *29*(1), 15-23.

Lele, P. P., Hosu, B. G., & Berg, H. C. (2013). Dynamics of mechanosensing in the bacterial flagellar motor. *Proceedings of the National Academy of Sciences*, *110*(29), 11839-11844.

Leone, V., & Faraldo-Gómez, J. D. (2016). Structure and mechanism of the ATP synthase membrane motor inferred from quantitative integrative modeling. *Journal of General Physiology*, 148(6), 441-457.

Leone, V., Pogoryelov, D., Meier, T., & Faraldo-Gómez, J. D. (2015). On the principle of ion selectivity in Na+/H+-coupled membrane proteins: experimental and theoretical studies of an ATP synthase rotor. *Proceedings of the National Academy of Sciences*, *112*(10), E1057-E1066.

Lorent, J. H., Levental, K. R., Ganesan, L., Rivera-Longsworth, G., Sezgin, E., Doktorova, M. D., ... & Levental, I. (2020). Plasma membranes are asymmetric in lipid unsaturation, packing and protein shape. *Nature Chemical Biology*, 1-9.

MacKenzie, K. R., Prestegard, J. H., & Engelman, D. M. (1997). A transmembrane helix dimer: structure and implications. *Science*, *276*(5309), 131-133.

Martin, J. L., Ishmukhametov, R., Spetzler, D., Hornung, T., & Frasch, W. D. (2018). Elastic coupling power stroke mechanism of the F1-ATPase molecular motor. *Proceedings of the National Academy of Sciences*, *115*(22), 5750-5755.

Meier, T., & Dimroth, P. (2002). Intersubunit bridging by Na+ ions as a rationale for the unusual stability of the c-rings of Na+-translocating F1F0 ATP synthases. *EMBO Reports*, *3*(11), 1094-1098.

Meier, T., Ferguson, S. A., Cook, G. M., Dimroth, P., & Vonck, J. (2006). Structural investigations of the membrane-embedded rotor ring of the F-ATPase from *Clostridium paradoxum*. *Journal of Bacteriology*, 188(22), 7759-7764.

Meier, T., Krah, A., Bond, P. J., Pogoryelov, D., Diederichs, K., & Faraldo-Gómez, J. D. (2009). Complete ion-coordination structure in the rotor ring of Na+-dependent F-ATP synthases. *Journal of Molecular Biology*, *391*(2), 498-507.

Meier, T., Morgner, N., Matthies, D., Pogoryelov, D., Keis, S., Cook, G. M., ... & Brutschy, B. (2007). A tridecameric c ring of the adenosine triphosphate (ATP) synthase from the thermoalkaliphilic *Bacillus sp. strain TA2.A1* facilitates ATP synthesis at low electrochemical proton potential. *Molecular Microbiology*, *65*(5), 1181-1192.

Meier, T., Polzer, P., Diederichs, K., Welte, W., & Dimroth, P. (2005). Structure of the rotor ring of F-Type Na+-ATPase from *Ilyobacter tartaricus*. *Science*, *308*(5722), 659-662.

Meier, T., Yu, J., Raschle, T., Henzen, F., Dimroth, P., & Muller, D. J. (2005). Structural evidence for a constant c11 ring stoichiometry in the sodium F-ATP synthase. *The FEBS Journal*, *272*(21), 5474-5483.

Mileykovskaya, E., & Dowhan, W. (2000). Visualization of phospholipid domains inEscherichia coli by using the cardiolipin-specific fluorescent dye 10-N-nonyl acridine orange. *Journal of Bacteriology*, *182*(4), 1172-1175.

Mitome, N., Suzuki, T., Hayashi, S., & Yoshida, M. (2004). Thermophilic ATP synthase has a decamer c-ring: indication of noninteger 10: 3 H+/ATP ratio and permissive elastic coupling. *Proceedings of the National Academy of Sciences*, 101(33), 12159-12164.

Morales-Rios, E., Montgomery, M. G., Leslie, A. G., & Walker, J. E. (2015). Structure of ATP synthase from *Paracoccus denitrificans* determined by X-ray crystallography at 4.0 Å resolution. *Proceedings of the National Academy of Sciences*, 112(43), 13231-13236.

Muench, S. P., Trinick, J., & Harrison, M. A. (2011). Structural divergence of the rotary ATPases. *Quarterly Reviews of Biophysics*, *44*(3), 311-356.

Mühleip, A. W., Joos, F., Wigge, C., Frangakis, A. S., Kühlbrandt, W., & Davies, K. M. (2016). Helical arrays of U-shaped ATP synthase dimers form tubular cristae in ciliate mitochondria. *Proceedings of the National Academy of Sciences*, *113*(30), 8442-8447.

Mühleip, A. W., Dewar, C. E., Schnaufer, A., Kühlbrandt, W., & Davies, K. M. (2017). In situ structure of trypanosomal ATP synthase dimer reveals a unique arrangement of catalytic subunits. *Proceedings of the National Academy of Sciences*, *114*(5), 992-997.

Mukherjee, S., & Warshel, A. (2012). Realistic simulations of the coupling between the protomotive force and the mechanical rotation of the F0-ATPase. *Proceedings of the National Academy of Sciences*, *109*(37), 14876-14881.

Mulkidjanian, A. Y., Makarova, K. S., Galperin, M. Y., & Koonin, E. V. (2007). Inventing the dynamo machine: the evolution of the F-type and V-type ATPases. *Nature Reviews Microbiology*, *5*(11), 892-899.

Mulkidjanian, A. Y., Galperin, M. Y., & Koonin, E. V. (2009). Co-evolution of primordial membranes and membrane proteins. *Trends in Biochemical Sciences*, 34(4), 206-215.

Müller, V., & Grüber, G. (2003). ATP synthases: structure, function and evolution of unique energy converters. *Cellular and Molecular Life Sciences CMLS*, *60*(3), 474-494.

Murphy, B. J., Klusch, N., Langer, J., Mills, D. J., Yildiz, Ö., & Kühlbrandt, W. (2019). Rotary substates of mitochondrial ATP synthase reveal the basis of flexible F1-Fo coupling. *Science*, *364*(6446), eaaw9128.

Nelson, N. (1992). Evolution of organellar proton-ATPases. *Biochimica et Biophysica Acta (BBA)-Bioenergetics*, *1100*(2), 109-124.

Nirody, J. A., Nord, A. L., & Berry, R. M. (2019). Load-dependent adaptation near zero load in the bacterial flagellar motor. *Journal of the Royal Society Interface*, *16*(159), 20190300.

Nishijima, S., Asami, Y., Uetake, N., Yamagoe, S., Ohta, A., & Shibuya, I. (1988). Disruption of the Escherichia coli cls gene responsible for cardiolipin synthesis. *Journal of Bacteriology*, *170*(2), 775-780.

Noji, H., Yasuda, R., Yoshida, M., & Kinosita, K. (1997). Direct observation of the rotation of F 1-ATPase. *Nature*, *386*(6622), 299-302.

Nord, A. L., Gachon, E., Perez-Carrasco, R., Nirody, J. A., Barducci, A., Berry, R. M., & Pedaci, F. (2017). Catch bond drives stator mechanosensitivity in the bacterial flagellar motor. *Proceedings of the National Academy of Sciences*, *114*(49), 12952-12957.

Oliver, P. M., Crooks, J. A., Leidl, M., Yoon, E. J., Saghatelian, A., & Weibel, D. B. (2014). Localization of anionic phospholipids in Escherichia coli cells. *Journal of Bacteriology*, *196*(19), 3386-3398.

Ortiz, A., Killian, J. A., Verkleij, A. J., & Wilschut, J. (1999). Membrane fusion and the lamellar-to-inverted-hexagonal phase transition in cardiolipin vesicle systems induced by divalent cations. *Biophysical Journal*, *77*(4), 2003-2014.

Oster, G., & Wang, H. (2000). Reverse engineering a protein: the mechanochemistry of ATP synthase. *Biochimica et Biophysica Acta (BBA)-Bioenergetics*, *1458*(2-3), 482-510.
25

Oster, G., Wang, H., & Grabe, M. (2000). How Fo–ATPase generates rotary torque. *Philosophical Transactions of the Royal Society of London. Series B: Biological Sciences*, *355*(1396), 523-528.

Paradies, G., Paradies, V., De Benedictis, V., Ruggiero, F. M., & Petrosillo, G. (2014). Functional role of cardiolipin in mitochondrial bioenergetics. *Biochimica et Biophysica Acta (BBA)-Bioenergetics*, *1837*(4), 408-417.

Paula, S., Volkov, A. G., Van Hoek, A. N., Haines, T. H., & Deamer, D. W. (1996). Permeation of protons, potassium ions, and small polar molecules through phospholipid bilayers as a function of membrane thickness. *Biophysical Journal*, 70(1), 339.

Paumard, P., Vaillier, J., Coulary, B., Schaeffer, J., Soubannier, V., Mueller, D. M., ... & Velours, J. (2002). The ATP synthase is involved in generating mitochondrial cristae morphology. *The EMBO Journal*, *21*(3), 221-230.

Penefsky, H. S., Pullman, M. E., Datta, A., & Racker, E. (1960). Partial resolution of the enzymes catalyzing oxidative phosphorylation II. Participation of a soluble adenosine triphosphatase in oxidative phosphorylation. *Journal of Biological Chemistry*, *235*(11), 3330-3336.

Pennington, E. R., Funai, K., Brown, D. A., & Shaikh, S. R. (2019). The role of cardiolipin concentration and acyl chain composition on mitochondrial inner membrane molecular organization and function. *Biochimica et Biophysica Acta (BBA)-Molecular and Cell Biology of Lipids*.

Pogoryelov, D., Klyszejko, A. L., Krasnoselska, G. O., Heller, E. M., Leone, V., Langer, J. D., ... & Meier, T. (2012). Engineering rotor ring stoichiometries in the ATP synthase. *Proceedings of the National Academy of Sciences*, *109*(25), E1599-E1608.

Pogoryelov, D., Reichen, C., Klyszejko, A. L., Brunisholz, R., Muller, D. J., Dimroth, P., & Meier, T. (2007). The oligomeric state of c rings from cyanobacterial F-ATP synthases varies from 13 to 15. *Journal of Bacteriology*, 189(16), 5895-5902.

Pogoryelov, D., Yu, J., Meier, T., Vonck, J., Dimroth, P., & Muller, D. J. (2005). The c15 ring of the Spirulina platensis F-ATP synthase: F1/F0 symmetry mismatch is not obligatory. *EMBO Reports*, *6*(11), 1040-1044.

Pogoryelov, D., Yildiz, Ö., Faraldo-Gómez, J. D., & Meier, T. (2009). High-resolution structure of the rotor ring of a proton-dependent ATP synthase. *Nature Structural & Molecular Biology*, *16*(10), 1068.

Preiss, L., Langer, J. D., Hicks, D. B., Liu, J., Yildiz, Ö., Krulwich, T. A., & Meier, T. (2014). The c-ring ion binding site of the ATP synthase from Bacillus pseudofirmus OF 4 is adapted to alkaliphilic lifestyle. *Molecular Microbiology*, *92*(5), 973-984.

Preiss, L., Langer, J. D., Yildiz, Ö., Eckhardt-Strelau, L., Guillemont, J. E., Koul, A., & Meier, T. (2015). Structure of the mycobacterial ATP synthase Fo rotor ring in complex with the anti-TB drug bedaquiline. *Science Advances*, 1(4), e1500106.

Preiss, L., Klyszejko, A. L., Hicks, D. B., Liu, J., Fackelmayer, O. J., Yildiz, Ö., ... & Meier, T. (2013). The c-ring stoichiometry of ATP synthase is adapted to cell physiological requirements of alkaliphilic *Bacillus pseudofirmus* OF4. *Proceedings of the National Academy of Sciences*, *110*(19), 7874-7879.

Preiss, L., Yildiz, Ö., Hicks, D. B., Krulwich, T. A., & Meier, T. (2010). A new type of proton coordination in an F1Fo-ATP synthase rotor ring. *PLoS Biology*, *8*(8).

Pullman, M. E., & Monroy, G. C. (1963). A naturally occurring inhibitor of mitochondrial adenosine triphosphatase. *Journal of Biological Chemistry*, *238*(11), 3762-3769.

Ramamurthi, K. S., Lecuyer, S., Stone, H. A., & Losick, R. (2009). Geometric cue for protein localization in a bacterium. *Science*, *323*(5919), 1354-1357.

Rand, D. M., Haney, R. A., & Fry, A. J. (2004). Cytonuclear coevolution: the genomics of cooperation. *Trends in Ecology & Evolution*, *19*(12), 645-653.

Ressl, S., van Scheltinga, A. C. T., Vonrhein, C., Ott, V., & Ziegler, C. (2009). Molecular basis of transport and regulation in the Na+/betaine symporter BetP. *Nature*, *458*(7234), 47-52.

Rexroth, S., zu Tittingdorf, J. M. M., Schwaßmann, H. J., Krause, F., Seelert, H., & Dencher, N. A. (2004). Dimeric H+-ATP synthase in the chloroplast of *Chlamydomonas reinhardtii*. *Biochimica et Biophysica Acta (BBA)-Bioenergetics*, *1658*(3), 202-211.

Reysenbach, A. L., & Shock, E. (2002). Merging genomes with geochemistry in hydrothermal ecosystems. *Science*, *296*(5570), 1077-1082.

Richter, M. L., Patrie, W. J., & McCarty, R. E. (1984). Preparation of the epsilon subunit and epsilon subunit-deficient chloroplast coupling factor 1 in reconstitutively active forms. *Journal of Biological Chemistry*, *259*(12), 7371-7373.

Romantsov, T., Battle, A. R., Hendel, J. L., Martinac, B., & Wood, J. M. (2010). Protein localization in Escherichia coli cells: comparison of the cytoplasmic membrane proteins ProP, LacY, ProW, AqpZ, MscS, and MscL. *Journal of Bacteriology*, *192*(4), 912-924.

Saddar, S., Dienhart, M. K., & Stuart, R. A. (2008). The F1F0-ATP synthase complex influences the assembly state of the cytochrome bc1-cytochrome oxidase supercomplex and its association with the TIM23 machinery. *Journal of Biological Chemistry*, *283*(11), 6677-6686.

Saffman, P. G., & Delbrück, M. (1975). Brownian motion in biological membranes. *Proceedings of the National Academy of Sciences*, *72*(8), 3111-3113.

Sakamoto, T., Inoue, T., Otomo, Y., Yokomori, N., Ohno, M., Arai, H., & Nakagawa, Y. (2012). Deficiency of cardiolipin synthase causes abnormal mitochondrial function and morphology in germ cells of *Caenorhabditis elegans*. *Journal of Biological Chemistry*, *287*(7), 4590-4601.

Schäfer, I., Rössle, M., Biuković, G., Müller, V., & Grüber, G. (2006). Structural and functional analysis of the coupling subunit F in solution and topological arrangement of the stalk domains of the methanogenic A1AO ATP synthase. *Journal of bioenergetics and biomembranes*, *38*(2), 83-92.

Schemidt, R. A., Qu, J., Williams, J. R., & Brusilow, W. S. (1998). Effects of carbon source on expression of Fo genes and on the stoichiometry of the c subunit in the F1Fo ATPase of *Escherichia coli*. *Journal of Bacteriology*, *180*(12), 3205-3208.

Schlegel, K., Leone, V., Faraldo-Gómez, J. D., & Müller, V. (2012). Promiscuous archaeal ATP synthase concurrently coupled to Na+ and H+ translocation. *Proceedings of the National Academy of Sciences*, *109*(3), 947-952.

Schulz, S., Iglesias-Cans, M., Krah, A., Yildiz, Ö., Leone, V., Matthies, D., ... & Meier, T. (2013). A new type of Na+-driven ATP synthase membrane rotor with a two-carboxylate ion-coupling motif. *PLoS Biology*, *11*(6).

Schulz, S., Wilkes, M., Mills, D. J., Kühlbrandt, W., & Meier, T. (2017). Molecular architecture of the N-type ATPase rotor ring from *Burkholderia pseudomallei*. *EMBO Reports*, *18*(4), 526-535.

Seelert, H., & Dencher, N. A. (2011). ATP synthase superassemblies in animals and plants: two or more are better. *Biochimica et Biophysica Acta (BBA)-Bioenergetics*, *1807*(9), 1185-1197.

Smith, J. B., Sternweis, P. C., & Heppel, L. A. (1975). Partial purification of active delta and epsilon subunits of the membrane ATPase from *Escherichia coli*. *Journal of Supramolecular Structure*, *3*(3), 248-255.

Sobti, M., Smits, C., Wong, A. S., Ishmukhametov, R., Stock, D., Sandin, S., & Stewart, A. G. (2016). Cryo-EM structures of the autoinhibited *E. coli* ATP synthase in three rotational states. *eLife*, *5*, e21598.

Song, J., Pfanner, N., & Becker, T. (2018). Assembling the mitochondrial ATP synthase. *Proceedings of the National Academy of Sciences*, 115(12), 2850-2852.

Steed, P. R., & Fillingame, R. H. (2008). Subunit a facilitates aqueous access to a membrane-embedded region of subunit c in Escherichia coli F1F0 ATP synthase. *Journal of Biological Chemistry*, *283*(18), 12365-12372.

Steed, P. R., & Fillingame, R. H. (2009). Aqueous accessibility to the transmembrane regions of subunit c of the *Escherichia coli* F1F0 ATP synthase. *Journal of Biological Chemistry*, *284*(35), 23243-23250.

Stock, D., Gibbons, C., Arechaga, I., Leslie, A. G., & Walker, J. E. (2000). The rotary mechanism of ATP synthase. *Current opinion in structural biology*, *10*(6), 672-679.

Strauss, M., Hofhaus, G., Schröder, R. R., & Kühlbrandt, W. (2008). Dimer ribbons of ATP synthase shape the inner mitochondrial membrane. *The EMBO Jurnal*, *27*(7), 1154-1160.

Sun, S. X., Wang, H., & Oster, G. (2004). Asymmetry in the F1-ATPase and its implications for the rotational cycle. *Biophysical Journal*, *86*(3), 1373-1384.

Symersky, J., Pagadala, V., Osowski, D., Krah, A., Meier, T., Faraldo-Gómez, J. D., & Mueller, D. M. (2012). Structure of the c 10 ring of the yeast mitochondrial ATP synthase in the open conformation. *Nature Structural & Molecular Biology*, *19*(5), 485.

Teixeira, F. K., Sanchez, C. G., Hurd, T. R., Seifert, J. R., Czech, B., Preall, J. B., ... & Lehmann, R. (2015). ATP synthase promotes germ cell differentiation independent of oxidative phosphorylation. *Nature Cell Biology*, *17*(5), 689-696.

van Lis, R., González-Halphen, D., & Atteia, A. (2005). Divergence of the mitochondrial electron transport chains from the green alga *Chlamydomonas reinhardtii* and its colorless close relative *Polytomella sp*. *Biochimica et Biophysica Acta (BBA)-Bioenergetics*, *1708*(1), 23-34.

van Lis, R., Mendoza-Hernández, G., Groth, G., & Atteia, A. (2007). New insights into the unique structure of the F0F1-ATP synthase from the chlamydomonad algae *Polytomella sp*. and *Chlamydomonas reinhardtii*. *Plant Physiology*, *144*(2), 1190-1199.

van Meer, G., Voelker, D. R., & Feigenson, G. W. (2008). Membrane lipids: where they are and how they behave. *Nature Reviews Molecular Cell Biology*, *9*(2), 112-124.

Vázquez-Acevedo, M., Cardol, P., Cano-Estrada, A., Lapaille, M., Remacle, C., & González-Halphen, D. (2006). The mitochondrial ATP synthase of chlorophycean algae contains eight subunits of unknown origin involved in the formation of an atypical stator-stalk and in the dimerization of the complex. *Journal of Bioenergetics and Biomembranes*, *38*(5-6), 271-282.

Vinothkumar, K. R., Montgomery, M. G., Liu, S., & Walker, J. E. (2016). Structure of the mitochondrial ATP synthase from *Pichia angusta* determined by electron cryo-microscopy. *Proceedings of the National Academy of Sciences*, 113(45), 12709-12714.

Vollmar, M., Schlieper, D., Winn, M., Büchner, C., & Groth, G. (2009). Structure of the c14 rotor ring of the proton translocating chloroplast ATP synthase. *Journal of Biological Chemistry*, *284*(27), 18228-18235.

von Ballmoos, C., Wiedenmann, A., & Dimroth, P. (2009). Essentials for ATP synthesis by F1F0 ATP synthases. *Annual Review of Biochemistry*, *78*, 649-672.

Wadhwa, N., Phillips, R., & Berg, H. C. (2019). Torque-dependent remodeling of the bacterial flagellar motor. *Proceedings of the National Academy of Sciences*, *116*(24), 11764-11769.

Walpole, T. B., Palmer, D. N., Jiang, H., Ding, S., Fearnley, I. M., & Walker, J. E. (2015). Conservation of complete trimethylation of lysine-43 in the rotor ring of c-subunits of metazoan adenosine triphosphate (ATP) synthases. *Molecular & Cellular Proteomics*, *14*(4), 828-840.

Wang, H., & Oster, G. (1998). Energy transduction in the F 1 motor of ATP synthase. *Nature*, *396*(6708), 279-282.

Watt, I. N., Montgomery, M. G., Runswick, M. J., Leslie, A. G., & Walker, J. E. (2010). Bioenergetic cost of making an adenosine triphosphate molecule in animal mitochondria. *Proceedings of the National Academy of Sciences*, *107*(39), 16823-16827.

Weber, J., & Senior, A. E. (2003). ATP synthesis driven by proton transport in F1F0-ATP synthase. *Febs Letters*, *545*(1), 61-70.

Wilkens, S., & Capaldi, R. A. (1998). ATP synthase's second stalk comes into focus. *Nature*, *393*(6680), 29-29.

Xing, J., Liao, J. C., & Oster, G. (2005). Making ATP. *Proceedings of the National Academy of Sciences*, *102*(46), 16539-16546.

Yamashita, A., Singh, S. K., Kawate, T., Jin, Y., & Gouaux, E. (2005). Crystal structure of a bacterial homologue of Na+/Cl--dependent neurotransmitter transporters. *Nature*, *437* (7056), 215-223.

Yanagisawa, S., & Frasch, W. D. (2017). Protonation-dependent stepped rotation of the F-type ATP synthase c-ring observed by single-molecule measurements. *Journal of Biological Chemistry*, *292*(41), 17093-17100.

Yasuda, R., Noji, H., Yoshida, M., Kinosita, K., & Itoh, H. (2001). Resolution of distinct rotational substeps by submillisecond kinetic analysis of F 1-ATPase. *Nature*, *410*(6831), 898-904.

Zhou, A., Rohou, A., Schep, D. G., Bason, J. V., Montgomery, M. G., Walker, J. E., ... & Rubinstein, J. L. (2015). Structure and conformational states of the bovine mitochondrial ATP synthase by cryo-EM. *eLife*, *4*, e10180.

Zhu, T. F., & Szostak, J. W. (2009). Coupled growth and division of model protocell membranes. *Journal of the American Chemical Society*, 131(15), 5705-5713.

Zimniak, L., Dittrich, P., Gogarten, J. P., Kibak, H., & Taiz, L. (1988). The cDNA sequence of the 69-kDa subunit of the carrot vacuolar H+-ATPase. Homology to the beta-chain of F0F1-ATPases. *Journal of Biological Chemistry*, *263*(19), 9102-9112.

zu Tittingdorf, J. M. M., Rexroth, S., Schäfer, E., Schlichting, R., Giersch, C., Dencher, N. A., & Seelert, H. (2004). The stoichiometry of the chloroplast ATP synthase oligomer III in *Chlamydomonas reinhardtii* is not affected by the metabolic state. *Biochimica et Biophysica Acta (BBA)-Bioenergetics*, 1659(1), 92-99.